\documentclass[openacc]{rstransa}

\usepackage{graphicx}
\usepackage{dcolumn}
\usepackage{bm,subfigure}
\usepackage{xcolor,comment}
\usepackage{hyperref}
\usepackage[mathlines]{lineno}

\titlehead{Research}

\begin{document}

\title{Interpretable and Equation-Free Response Theory for Complex Systems}

\author{Valerio Lucarini$^{1}$}

\address{$^{1}$School of Computing and Mathematical Sciences, University of Leicester, Leicester, LE17RH UK}

\subject{Applied Mathematics and Theoretical Physics, Statistical Physics}

\keywords{Markov chains; Response Theory; Koopmanism; Multiple Time Scales; Model Reduction; Markov State Modelling; Prony Method}

\corres{Valerio Lucarini\\
\email{v.lucarini@leicester.ac.uk}}

\begin{abstract}
Response theory provides a pathway for understanding the sensitivity of a system and for predicting how its statistical properties change when a perturbation is applied. In the case of complex and multiscale systems, to achieve enhanced practical applicability, response theory should be interpretable, capable of focusing on relevant timescales, and amenable to data-driven and equation-agnostic implementations. Along these lines, in the spirit of Markov state modelling, we present linear and nonlinear response formulas for Markov chains. We obtain simple and easily implementable expressions that can be used to predict the response of observables as well as of higher-order correlations. The methodology proposed here can be implemented in a purely data-driven setting and even if the underlying evolution equations are unknown. The use of algebraic expansions inspired by Koopmanism allows to elucidate the role of different time scales and modes of variability, and to find explicit and interpretable expressions for the Green's functions at all orders. This is a major advantage of the framework proposed here. We illustrate our methodology in a very simple yet instructive metastable system. Finally, our results provide a dynamical foundation for the Prony method, which is commonly used for the statistical analysis of discrete time signals.\end{abstract}

\begin{fmtext}

\end{fmtext}

\maketitle
\section{Introduction}\label{introduction}
Response theory in statistical mechanics constitutes a powerful framework for analyzing the behavior of a large variety of systems subjected to external perturbations. It provides a powerful and unifying paradigm for connecting the microscopic dynamics and reference statistical properties of a system to its macroscopic response under external influences. Its foundations have been widely discussed in the mathematical literature and its applications permeate various domains of physics, chemistry, biology,  materials science, and quantitative social sciences \cite{Hanggi1982,Baiesi2013,Sarracino2019}. 
At the core of response theory lies the formulation of response functions, which quantify the system's reaction to acting forcings. In the case of systems at thermodynamic equilibrium and considering for the moment only the linear approximation to the response, the fluctuation-dissipation theorem (FDT) establishes that such functions are expressed in terms of time-lagged correlation between suitably defined observables on the unperturbed state \cite{Kubo1966}. 

However, response theory is not limited to equilibrium systems \cite{Hairer2010}. For nonequilibrium systems possessing smooth invariant measure with respect to Lebesgue, as in the case of stochastic dynamical systems forced by sufficiently non-degenerate noise, one can still express the response formulas in terms of time-lagged correlations, which implies the existence of a clear correspondence between forced and free fluctuations of the system \cite{marconi2008fluctuation,pavliotisbook2014}. Things become more problematic in the case of dissipative chaotic systems. Here, as a result of the singularity of the invariant measure with respect to Lebesgue, the FDT does not apply and there is no full equivalence between forced and free fluctuations. Yet, making suitable assumptions on the dynamics, it is possible to establish a response theory also in this case. The original results proposed by Ruelle, which required fairly restrictive conditions of uniform hyperbolicity \cite{ruellegeneral1998,ruelle2009}, have then been clarified and extended using functional analytical techniques  \cite{Liverani2006,Baladi2014,Baladi2017a}.

When the perturbation is large or when the system exhibits a very amplified response, the linear approximation linking the amplitudes of the forcing and of the response breaks down, so that higher-order terms in the system's response need to be considered \cite{ruelle_nonequilibrium_1998,lucarini2008,Lucarini2009DispersionNonLinearResponseLorenz}. Nonlinear response operators have a  convoluted structure and depend on multiple time variables. The nonlinear response describes  more complex interplay between internal feedbacks and acting forcings and, if more than one forcing is present, accounts for the interplay - which can be synergistic or antagonistic - of the various acting forcings. Nonlinear effects become quantitatively dominant in the proximity of critical transitions, which are associated with the divergence of the response of the system. However, the occurrence of such divergent behavior can be captured simply by looking at the linear response of the system \cite{Chekroun_al_RP2,Santos2022,LC2023}.

A key difficulty of response formulas is that they are based on expressions that do not provide a clear imprint of the dominant modes of variability of the system. A way forward in this direction is provided by  Koopmanism \cite{Mezic2005,Budisic2012,Kutz2016}, which, roughly speaking, transforms nonlinear dynamics into a linear {\color{black}dynamical} framework in an infinite dimensional space of observables, where the {\color{black}key} information is contained in the  eigenfunction and modes of the Koopman (or Kolmogorov in the case of stochastic dynamics) operator. From now on, with an abuse of language, we will {\color{black}use the expression \textit{Koopman operator}} also in the stochastic case.  In practice, one needs to approximate the Koopman operator using data-driven methods such as the extended dynamic mode decomposition (eDMD) \cite{brunton2022data}. One considers
a finite dictionary of observables, and each eigenvector of the finite-dimensional approximation of the Koopman operator is  obtained as a linear combination of such observables \cite{Klus2018}.  See \cite{Colbrook2024Multi} for a comprehensive review of eDMD methods

By linking response theory with Koopmanism, one can derive interpretable representations of the system’s response to perturbations. The use of Koopmanism enables the decomposition of the response into contributions from distinct  modes of natural variability of the system \cite{Santos2022,LC2023}.  Recently, we have been successful in merging algorithmically response theory and Koopmanism \cite{Zagli2024,Lucarinietal2025} and in showing that Koopmanism provides a pathway for extending response theory to the very relevant yet so far unexplored case where the stochasticity includes jump processes \cite{Chekroun2024Kolmogorov}. While these  results are extremely encouraging, a nontrivial hurdle that still needs to be overcome is the applicability of this methodology to high-dimensional systems. 

.\subsection{A Pipeline for an Interpretable and Equation-free Response Theory}\label{pipeline}

A finite-state, discrete-time Markov chain is a stochastic process that describes the sequence of possible events, chosen among a finite set, evolving, at discrete time, according to a specified probabilistic rule. The probability of occurrence of the future state depends only on the current state and not on past states.  Markov chains can be associated with directed graphs, whereby the states correspond to the nodes of the network, and the entries of the Markov matrix give the  weight of the directed edge between two nodes \cite{Norris1998,behrends2000introduction}. Markov chains are extremely relevant to understand the properties of general dynamical systems \cite{bowen1970markov,attal2010markov}. The so-called Ulam method \cite{Ulam1960} approximates the Perron-Frobenius operator, which pushes forward the probability distributions of a dynamical system \cite{baladi2000positive}, by a Markov matrix whose entries  represent transition probabilities between partition elements occurring for finite time horizons. The Ulam Conjecture  states that {\color{black}the invariant measure of the finite Markov matrix converges to that of the true dynamical system as one considers finer and finer partitions} \cite{Froyland1998}. 

While the Ulam method is by itself essentially a brute-force approximation, and its convergence is usually slow \cite{Ding2002}, it is possible to use it very effectively. Specifically, Markov state modeling (MSM) is a smart Ulam method that is particularly effective for studying systems with complex dynamics that evolve across multiple timescales. In this framework, the continuous phase space of a system is optimally discretized into a finite number of so-called (micro)states. Each state corresponds to a cell of the Vorono\"i tessellation \cite{Aurenhammer1991} constructed following k-means clustering \cite{Forgy65,Lloyd82} of the data.
The transitions between such states are governed by a Markov chain \cite{Laio2006,Pande2010,Bowman2014,Husic2018}, which contains all the information needed to describe the statistics and dynamics of the system at the coarse-grained level. In this case, the slowest time scales are associated with the relaxation between the main metastable states, and {\color{black}metastable regions can be identified by studying the level sets of the dominant modes of the Koopman operator \cite{Froyland2014}. Instead, the dynamical processes occurring within each metastable region are associated with faster timescales}. A maximally reduced version of MSM targets directly the metastable states and studies exclusively the transition rates between such states \cite{Bittracher2018}.

{\color{black} The Mori-Zwanzig theory \cite{mori_transport_1965,zwanzig_memory_1961} has emerged as possibly the key statistical mechanical paradigm for  extracting accurate coarse-grained models from multiscale systems; see discussion in e.g. \cite{Kalliadasis2015,Chekroun2015b,santos2021reduced,Chekroun2021,Chekroun2025}. Nonetheless, it has recently become apparent that one might desire to cast the problem of deriving the coarse-grained model not necessarily as an approximate or exact analytical exercise, but rather in purely algorithmic terms \cite{Gear2003,Kevrekidis2009}. 

Along these lines, }MSM has the great advantage of being a) equation-agnostic: it is a data-driven method that can be deployed on observed or modelling data and is oblivious to the underlying evolution equations; {\color{black}b) constructed in such a way that memory effects are effectively neglected; and c)} (possibly) able to beat the curse of dimensionality,  
because the geometric and dynamical complexity of the original system is bypassed once one is able to define smartly the basis of states associated with the markovian dynamics. {\color{black}Points b) and c) are facilitated if one is able to define reaction coordinates, i.e. collective variables able to provide a low dimensional description of the key macroscopic features of the system. Defining reaction coordinates for general complex systems that are far from equilibrium is highly nontrivial, as symmetry or thermodynamic arguments used for equilibrium dynamics cannot be simply adapted \cite{Ma2005,Laio2006,Rogal2021}; see discussion in \cite{ZagliLucariniPavliotis,ZaglietalJPA2024}. } Applying the Ulam method to a dynamical systems amounts to considering a Koopman  dictionary {\color{black}comprising of the characteristic function of all cells of the tessellation. Whilst considering a dictionary of discontinuous functions can cause headaches, the use of characteristic functions has also some clear advantages;} see at this regard the recently proposed multiplicative DMD algorithm \cite{Boulle2024}.


Comprehensive and easily implementable response formulas for Markov chains are - apart from their intrinsic interest - of great practical utility in the analysis of a complex system because they can be directly applied to its coarse-grained representation constructed according to MSM protocols and thus bypassing (and obliviously to) the underlying evolution equations. 

We  discussed elsewhere  how the invariant measure of a Markov chain responds to a time-independent perturbation, presenting explicit bounds for the validity of classic perturbative approach and providing explicit formulas for linear and nonlinear terms, including renormalized results  \cite{Lucarini2016,SantosJSP}. Independent results that delve more deeply in the physical interpretation of the response formulas have been recently reported \cite{Esposito2024a,EspositoPRL2024b}.

Treating carefully the transition from the microscopic description of a system to its heavily coarse-grained representation as a discrete Markov process and linking the properties of fluctuations and response across scales is an extremely challenging task, see \cite{Zhang2012,Koltai2018,Falasco2025}. Here we set ourselves in a much simplified  setting and we assume that upstream of our work someone has carefully constructed a coarse-grained representation of the system as a discrete Markov chain, e.g. by applying MSM to a system in a reference state and in a slightly perturbed state. Both states are characterised by autonomous dynamics. Our goal is to predict how different time modulations of the forcing  impacts the statistical properties of the coarse-grained system. Hence, by construction, we will neglect the subscale processes. 

 We will derive   formulas that are able to predict the linear as well as higher order response of the coarse-grained system for general, time-dependent perturbation via simple matrix relations. Our ability to treat explicitly time-independent perturbation is, as far as we know, novel, and goes in the direction of analysis of entropy production for non autonomous systems \cite{Seifert2005}. Response formulas can be derived for observables as well as for lagged correlations between observables, thus allowing for predicting how the forcing impacts the variability of the system. The latter had been attempted in a previous work but only in the case of static forcings \cite{LucariniWouters2017}. We will also provide a simple but possibly very instructive novel way of expressing the linear and nonlinear response operators for Markov chains by  taking advantage of the Koopman formalism for finite-state processes that clarifies the roles of the time scales that are intrinsic to the system. 
 
 The rest of the paper is structured as follows.
 The derivation and discussion of  response formulas is presented in Sect. \ref{response} for observables in in Sect. \ref{responsecorrelations} for correlations.
  In order to illustrate some of our findings, we will provide in Sect. \ref{model} a proof-of-concept application of some of our key results on a simple yet instructive two-dimensional (2D) Langevin equation closely related to an example provided in \cite{Klus2016} which is characterized by nontrivial metastability properties. 
In Sect. \ref{conclusions} we present a discussion of our results, including a comment on their implications for providing a dynamical foundation for the time-series analysis Prony method \cite{Hua1990,Park1999,Kunis2016,Rodriguez2018}, as well as perspectives for future investigations. Additionally App. \ref{arbitrary}, App. \ref{responsedynamiccorrelation}, and App. \ref{ep} provide general formulas for the nonlinear response of observables, the linear response of correlation functions, and the linear response of entropy production to time-dependent perturbations, respectively.

\section{Response Theory for Markov Chains: Time-Dependent Perturbations}\label{response}
We set ourselves in the same framework described in \cite{Lucarini2016,SantosJSP}. Let us consider a mixing $N$-state ($N$ is finite) Markov process defined by the  matrix $\mathcal{M}\in\mathbb{R}^{N\times N}$. $\mathcal{M}_{ij}\geq0$ is a stochastic matrix that measures the probability of reaching the state $i$ at time $n$ given that at time $n-1$ the system is in the state $j$. Since the process is mixing we can reach any state $i$ starting from any state $j$ is we wait a sufficiently long time, or, more specifically $\exists p\geq 1 | \mathcal{M}^p_{ij}>0$. We consider the eigenvalue problem $\mathcal{M}\mathbf{v}=\lambda\mathbf{v}$. For the Perron-Frobenius theorem, there is a unique invariant measure, i.e. $\exists! {\nu_{inv}}|\mathcal{M} {\nu_{inv}}={\nu_{inv}}$, so that ${\nu_{inv}}\in\mathbb{R}^{N\times 1}$ defines the invariant measure associated with unitary eigenvalue \cite{senetanonnegative1973}. We also have pairs $\{\lambda_j,{\nu_j}\}$, such that $\mathcal{M} {\nu_j}=\lambda_j{\nu_j}$, with $|\lambda_j|<1$ and $\nu_j\in\mathbb{R}^{N\times 1}$ for $j=2,\ldots N$. {\color{black}Customarily, we label the eigenvalues in such a way that they are ordered by their magnitude, i.e. $|\lambda_i|\geq |\lambda_j|$ if $i<j$.} |Additionally, we have that $\sum_{i=1}^N (\nu_{inv})_i=1$ and $\sum_{i=1}^N (\nu_{j})_i=0$, $j>1$. 

Let us now consider a  perturbation of the form $\mathcal{M}\rightarrow \mathcal{M}_{\epsilon,n}=\mathcal{M} +\epsilon f(n) m$, where $f:\mathbb{N}\rightarrow\mathbb{R}$  defines a time-dependent modulation with $|f(n)|<1$, $m\in \mathbb{R}^{N\times N}$ , and $\epsilon$ is a small real number.  We impose that $\mathcal{M}_{\epsilon,n}$ is at all times a stochastic matrix. Hence, $\sum_i m_{ij} =0$. The perturbed Markov chain evolves according to the following law:

\begin{align}
    \nu(n)= \mathcal{M}_{\epsilon,n}\nu(n-1)= (\mathcal{M} +\epsilon f(n-1) m )\nu(n-1)
\end{align}
We plug $\nu(n)=\nu_{inv}+\epsilon\nu^{(1)}(n) + h.o.t.$ in the equation above and collect the terms proportional to $\epsilon$. The conditions behind the applicability of the perturbative approach are discussed in detail in   \cite{Lucarini2016,SantosJSP} and will not be repeated here; see also \cite{Abbas2016} and recent review devoted to continuous time Markov chains  \cite{Mitrophanov_2024}. It suffices here to say that {\color{black}if the operator $\mathcal{M}$ has a finite spectral gap, we can define an $\epsilon_{max}>0$ such that $\forall \epsilon$ with $|\epsilon|<\epsilon_{max}$ the perturbative expansions converge.} We obtain:
\begin{align}
    \nu^{(1)}(n) &= \mathcal{M}\nu^{(1)}(n-1)  + f(n-1) m \nu_{inv}
\end{align}
By applying recursively the relationship above and considering that $\lim_{n\rightarrow\infty} |\mathcal{M}^n \nu^{(1)}(n)|=0$, we have:
\begin{align}\label{deltanu1}
    \nu^{(1)}(n) &= \sum_{k=0}^\infty \mathcal{M}^k  m \nu_{inv}f(n-k-1)\\
    &=\sum_{k=-\infty}^\infty \Theta(k)  \mathcal{M}^k  m \nu_{inv}f(n-k-1)
\end{align}
where $\Theta(k)=1$ if $k\geq0$ and $\Theta(k)=0$ if $k<0$. 
\subsection{Linear Response}
Let us define $\langle \Psi,\mu \rangle=\sum_{i=1}^N \Psi_i \mu_i$ the the expectation value of an observable $\Psi\in\mathbb{R}^{1\times N}$ according to the measure $\mu$. We then have $\langle \Psi, \nu(n) \rangle=\langle \Psi,{\nu_{inv}}\rangle +\epsilon \langle \Psi,\nu^{(1)}_i(n) \rangle+h.o.t.$. We then have:
\begin{align}\label{Greenm}
    \frac{\mathrm{d}\langle \Psi,\nu(n) \rangle}{\mathrm{d}\epsilon}\big{|}_{\epsilon=0}&=\langle \Psi,\nu^{(1)}(n) \rangle\nonumber \\
    &= \sum_{k=-\infty}^\infty \Theta(k) \langle m^T  (\mathcal{M}^T)^k  \Psi,\nu_{inv}\rangle f(n-k-1)\nonumber \\
    & =  (\mathcal{G}^{(1)}_{m,\Psi}\star f)(n), \quad \mathcal{G}^{(1)}_{m,\Psi}(k)=\Theta(k) \langle m^T  (\mathcal{M}^T)^k  \Psi,\nu_{inv}\rangle     \end{align}
where $\mathcal{G}^{(1)}_{m,\Psi}(k)$ is the (causal) first order Green's function {\color{black} and $\star$ indicates the convolution product}.  We define the Koopman operator $\mathcal{K}=\mathcal{M}^T$. {\color{black}We assume the absence of degeneracies and we define $\Lambda\in\mathbb{R}^{N\times N}=diag(\lambda_1,\ldots,\lambda_N)$. 
Hence, $\mathcal{K}=V\Lambda V^{-1}$, with $V\in\mathbb{R}^{N\times N}$.} We have that 
\begin{align}
\mathcal{K}^m=\sum_{i=1}^N\lambda^m_iv_iw_i^T = \sum_{i=1}^N\lambda^m_i \Pi_i 
\end{align}
where $v_i$ is the $i^{th}$ right eigenvector, $w_i$ is the  $i^{th}$ left eigenvector of $K$ and $\Pi_i$ defines the projector on the $i^{th}$ eigenmode of $\mathcal{K}$.\footnote{Clearly, we have that $\mathcal{M}^m=\sum_{i=1}^N\lambda^m_iw_iv_i^T = \sum_{i=1}^N\lambda^m_i \mathcal{Q}_i$, where $\mathcal{Q}_i=\Pi_i^T$ is the projector on the $i^{th}$ eigenmode of the Perron-Frobenius operator.}
By inserting the previous expression in the definition of the Green's function we obtain:
\begin{align}
\mathcal{G}^{(1)}_{m,\Psi}(k)&=\Theta(k) \langle m^T  \sum_{i=2}^N\lambda^k_iv_iw_i^T  \Psi,\nu_{inv}\rangle \nonumber\\
&=\Theta(k) \langle m^T  \sum_{i=2}^N\lambda^k_i\Pi_i \Psi,\nu_{inv}\rangle=\sum_{i=1}^N \mathcal{G}^{(1)}_{m,\Psi,i}(k)= \sum_{i=2}^N \mathcal{G}^{(1)}_{m,\Psi,i}(k)\label{Gdecompo}
\end{align}
where 
\begin{equation}\label{KoopmanResponseb}
\mathcal{G}^{(1)}_{m,\Psi,i}(k)= \Theta(k) \alpha_i\lambda_i^k \quad \alpha_i= \langle m^T  \Pi_i \Psi,\nu_{inv}\rangle
\end{equation}
where we have broken down the Green's function into the sum of $N-1$ terms, each associated with a specific mode of variability of the system. Note that the first term $i=1$ in the summation given in Eq. \ref{Gdecompo} vanishes because it can be proved that $\Pi^T_1 m=0$ \cite{Lucarini2016}. 
Since {\color{black}$\lambda_i^k=\exp(k\beta_i)$, with $\Re[\beta_i]<0$}, the previous expansion provides a specific statistical model - the sum of exponentials - for fitting a Green's function from data. We will comment on this matter in Sect. \ref{conclusions}. The results presented in Eqs. \ref{Gdecompo}-\ref{KoopmanResponseb} correspond, in the case of a finite-state Markov chain, to the key findings shown in \cite{Santos2022} for a general Langevin equation and a general Koopman  dictionary. The derivation is much simpler in the case presented here whilst very little is lost at conceptual level, as we are {\color{black}indeed considering, as discussed in Subsect. \ref{introduction}\ref{pipeline}}, the case where the dictionary is given by the characteristic functions of the elements of the tessellation in phase space defining the states of Markov chain. 

{\color{black}Finally,} by inserting the previous expression in the linear response formula above and by rearranging terms, we have:
\begin{equation}
    \frac{\mathrm{d}\langle \Psi,\nu(n) \rangle}{\mathrm{d}\epsilon}\big{|}_{\epsilon=0} = \sum_{i=1}^N (\mathcal{G}^{(1)}_{m,\Psi,i}\star f)(n),   
\end{equation}
which separates the linear response formula into $N$ distinct contributions. Each of this contributions can be computed from the knowledge of $m$, $\mathcal{M}$, $\Psi$, and $f$. 
\subsection{Second order response}\label{second}
Let's now consider the full perturbative expansion  $\nu(n)=\nu_{inv}+\sum_{l=1}^\infty \epsilon^l\nu^{(l)}(n)$. By equating terms proportional to powers of $\epsilon$ larger than one, we obtain::
\begin{align}
    \nu^{(l)}(n) &= \mathcal{M}\nu^{(l)}(n-1)  + f(n-1) m \nu^{(l-1)}(n),\quad l>1 
\end{align}
By applying recursively over times the relationship above and considering that $\lim_{n\rightarrow\infty} |\mathcal{M}^n \nu^{l1)}(n)|=0$, we have:
\begin{align}
    \nu^{(l)}(n) &= \sum_{k=0}^\infty \mathcal{M}^k  m \nu^{(l-1)}(n-k-1)f(n-k-1)\\
    &=\sum_{k=-\infty}^\infty \Theta(k)  \mathcal{M}^k  m \nu^{(l-1)}(n-k-1)f(n-k-1)
\end{align}
Let's now consider the second order term $\nu^{(2)}(n)$. We  have 
\begin{align}
    \nu^{(2)}(n) &= \sum_{k=0}^\infty \mathcal{M}^k  m \nu^{(1)}(n-k-1)f(n-k-1)\\
    &=\sum_{k=-\infty}^\infty \sum_{p=-\infty}^\infty\Theta(k)\Theta(p)  \mathcal{M}^k  m    \mathcal{M}^p  m \nu_{inv}f(n-k-p-2)f(n-k-1)
\end{align}
where we have used the expression for $\nu^{(1)}$ above. Note that the $\Theta$'s ensure a {\color{black}correctly time-ordered consideration of the acting perturbation.} 

Let's now consider the second order response of a generic observable $\Psi$. We have:
\begin{align}
    \frac{1}{2}&\frac{\mathrm{d^2}\langle \Psi,\nu(n) \rangle}{\mathrm{d}\epsilon^2}\big{|}_{\epsilon=0}=\langle \Psi,\nu^{(2)}(n) \rangle\\
    &= \sum_{k=-\infty}^\infty \Theta(k) \langle m^T  (\mathcal{M}^T)^k  \Psi,\nu^{(1)}(n-k-1)\rangle f(n-k-1)\\
    &=\sum_{k=-\infty}^\infty \sum_{p=-\infty}^\infty\Theta(k)\Theta(p)  \langle m^T(\mathcal{M}^T)^p m^T(\mathcal{M}^T)^k \Psi ,\nu_{inv}\rangle f(n-k-p-2)f(n-k-1)\\
    & =  (\mathcal{G}^{(2)}_{m,\Psi}\star f)(n)  
\end{align}
where $\mathcal{G}^{(2)}_{m,\Psi}(k,p)=\Theta(k)\Theta(p)\langle m^T(\mathcal{M}^T)^p m^T(\mathcal{M}^T)^k \Psi ,\nu_{inv}\rangle$ and where $*$ indicates here a double convolution sum.
By  using $\mathcal{K}^m= (\mathcal{M}^T)^m=\sum_{i=1}^N\lambda^m_i \Pi_i$ in the expression of the second order Green's function above, one can express it as a double sum of terms, each describing the contribution to the nonlinear response coming from a specific pair of Koopman modes. Indeed, we have:
\begin{align}\label{decompositionG2}
    \mathcal{G}^{(2)}_{m,\Psi}(k,p)=\Theta(k)\Theta(p)\sum_{i,j=2}^N\alpha_{ij}\lambda_i^k\lambda_j^p, \quad \alpha_{ij}=\langle m^T \Pi_j m^T\Pi_i \Psi ,\nu_{inv}\rangle
\end{align}
Note again that our summation excludes the term corresponding to the invariant measure of the system. The results above can easily be extended at all orders of perturbations, see Appendix \ref{arbitrary}. This implies that the full nonlinear time-dependent response can be obtained from the knowledge of $m$, $\mathcal{M}$, and $f$ for any observable of the system. Assuming $f=1$ and taking the $n\rightarrow\infty$ limit, the results presented in \cite{Lucarini2016} can be easily recovered. 

\section{Response theory for Correlations}\label{responsecorrelations}
In the vast majority of cases, response theory has been used to  computate the change of the measure of a system resulting from applied forcings. Yet, in many practical cases, it is relevant to study the impact of the perturbation on the temporal correlation properties of the system. Taking the example of climate science to illustrate this point, response theory applied to observables describes the change in the state of the climate at a certain time horizon with respect to a reference climatology, whilst response theory applied to correlation describes how climate variability and so-called teleconnections like the North Atlantic Oscillation or El-Ni\~no-Southern Oscillation are impacted by the applied perturbation \cite{GhilLucarini2020,LucariniChekroun2023}. 
Some preliminary contributions to the development of response formulas for correlations have been presented in \cite{LucariniWouters2017}, which provides a reference for the results presented below. 

Let us define $C_l(\Psi,\Phi)=\langle (\mathcal{M}^T)^l\Psi\circ \Phi,\nu_{inv} \rangle-\langle \Psi,\nu_{inv}\rangle\langle\Phi,\nu_{inv}\rangle$  the unperturbed  $l-$lagged correlation between the two observables $\Psi$ and $\Phi$, where $\circ$ indicates the Hadamard product. 
We first consider the case of a static perturbation to the Markov process of the form $\mathcal M\rightarrow \mathcal{M}_{\epsilon}=\mathcal{M}+\epsilon m$, which has not yet been explicitly treated in the literature up to our knowledge. We then have 
$C^\epsilon_l(\Psi,\Phi)=\langle (\mathcal{M}^T+\epsilon m^T)^l\Psi \circ \Phi,\nu_{\epsilon} \rangle-\langle \Psi,\nu_{\epsilon}\rangle\langle\Phi,\nu_{\epsilon}\rangle$. Now we write down the terms up to first order in $\epsilon$ in the expression of the correlation. For the first term, we obtain:
\begin{align}\label{corre1}
     &\langle (\mathcal{M}^T+\epsilon m^T)^l\Psi \circ \Phi,\nu_{inv}+\sum_{p=1}^\infty\epsilon^p\nu^{(p)}\rangle = \langle ( \mathcal{M}^T)^l\Psi \circ \Phi,\nu_{inv}\rangle  \nonumber \\
     & + \epsilon  \sum_{q=0}^{l-1} \langle (\mathcal{M}^{l-q-1})^T m^T (\mathcal{M}^{q})^T\Psi\circ\Phi,\nu_{inv}\rangle + \epsilon \sum_{k=0}^\infty \langle (\mathcal{M}^T)^l\Psi \circ \Phi,\mathcal{M}^k m \nu_{inv}\rangle  +  o(\epsilon)
\end{align}
For the second term, we have
\begin{align}\label{corre2}
\langle \Psi,\nu_{inv}+\sum_{p=1}^\infty\epsilon^p\nu^{(p)}\rangle &\langle\Phi,\nu_{inv}+\sum_{p=1}^\infty\epsilon^p\nu^{(p)}\rangle \rangle=\langle \Psi,\nu_{inv}\rangle \langle\Phi,\nu_{inv}\rangle\nonumber\\
&\epsilon \sum_{k=0}^\infty \langle \Psi, \mathcal{M}^k m \nu_{inv} \rangle\langle\Phi,\nu_{inv}\rangle+\epsilon \langle\Psi,\nu_{inv}\rangle\sum_{k=0}^\infty \langle \Phi, \mathcal{M}^k m \nu_{inv} \rangle +o(\epsilon).
\end{align}
Keeping in mind that $\sum_{k=0}^{\infty}\mathcal{M}^k=(1-\mathcal{M}+\mathcal{Q}_1)^{-1}=\mathcal{Z}$, we obtain:
\begin{align}\label{decomporesponse}
&\frac{\mathrm{d}C^\epsilon_l(\Psi,\Phi)}{\mathrm{d}\epsilon}|_{\epsilon=0}= \underbrace{\sum_{q=0}^{l-1} \langle (\mathcal{M}^{l-q-1})^T m^T (\mathcal{M}^{q})^T\Psi\circ\Phi,\nu_{inv}\rangle}_{\delta^{(1)}_{a,\epsilon} (\Psi(l),\Phi)}\nonumber \\
&+\underbrace{\langle m^T\mathcal{Z}^T (\mathcal{M}^T)^l\Psi \circ \Phi, \nu_{inv}\rangle}_{\delta^{(1)}_{b,\epsilon} (\Psi(l),\Phi)}
-\underbrace{\langle\Phi,\nu_{inv}\rangle\langle m^T\mathcal{Z}^T \Psi,\nu_{inv}\rangle}_{\langle \Phi \rangle|_{\epsilon=0} \partial \langle \Psi \rangle/\partial\epsilon |_{\epsilon=0}}
-\underbrace{\langle\Psi,\nu_{inv}\rangle\langle m^T\mathcal{Z}^T \Phi,\nu_{inv}\rangle}_{\langle \Psi \rangle|_{\epsilon=0} \partial \langle \Phi \rangle/\partial\epsilon |_{\epsilon=0}}
\end{align}
Hence, the sensitivity of a time lagged correlation between $\Psi$ and $\Phi$ is nontrivial and can be broken up into four terms. The first term $\delta^{(1)}_{a,\epsilon} (\Psi(l),\Phi)$ is associated with the impact of perturbation on the evolution law of the system in the time interval of length $l$. The second term $\delta^{(1)}_{b,\epsilon} (\Psi(l),\Phi)$ describes the change in the expectation value of the product $\Psi(l),\Phi$ due to the variation of the invariant measure. The last two terms associated with the change in the expectation value of the two observables. Note that the matrix expression provided here is much simpler than the corresponding functional expression given in \cite{LucariniWouters2017}. We remark that by performing the spectral expansion of the $\mathcal{M}^T$ matrix we are able to disentangle the contributions coming from the various modes of the Koopman operator. 

It is indeed possible to extend the response theory for correlation to the case where $f$ has a non-trivial time dependence. The results are reported in Appendix \ref{responsedynamiccorrelation}. It is important to note that the resulting formulas allow us to define how correlations behave in a non-autonomous system. In this case, correlations at time $t$ need to be interpreted as integrals performed on the measure of the snapshot attractor at time $t$ (which is a slice of the pullback attractor \cite{Ghil2008,Chekroun2011}); see discussion in \cite{bodai2011,Tel2020,Bodai2020}. 

\section{A Simple Example}\label{model}
\begin{figure}
    \centering  a)\subfigure{\includegraphics[width=0.47\textwidth]{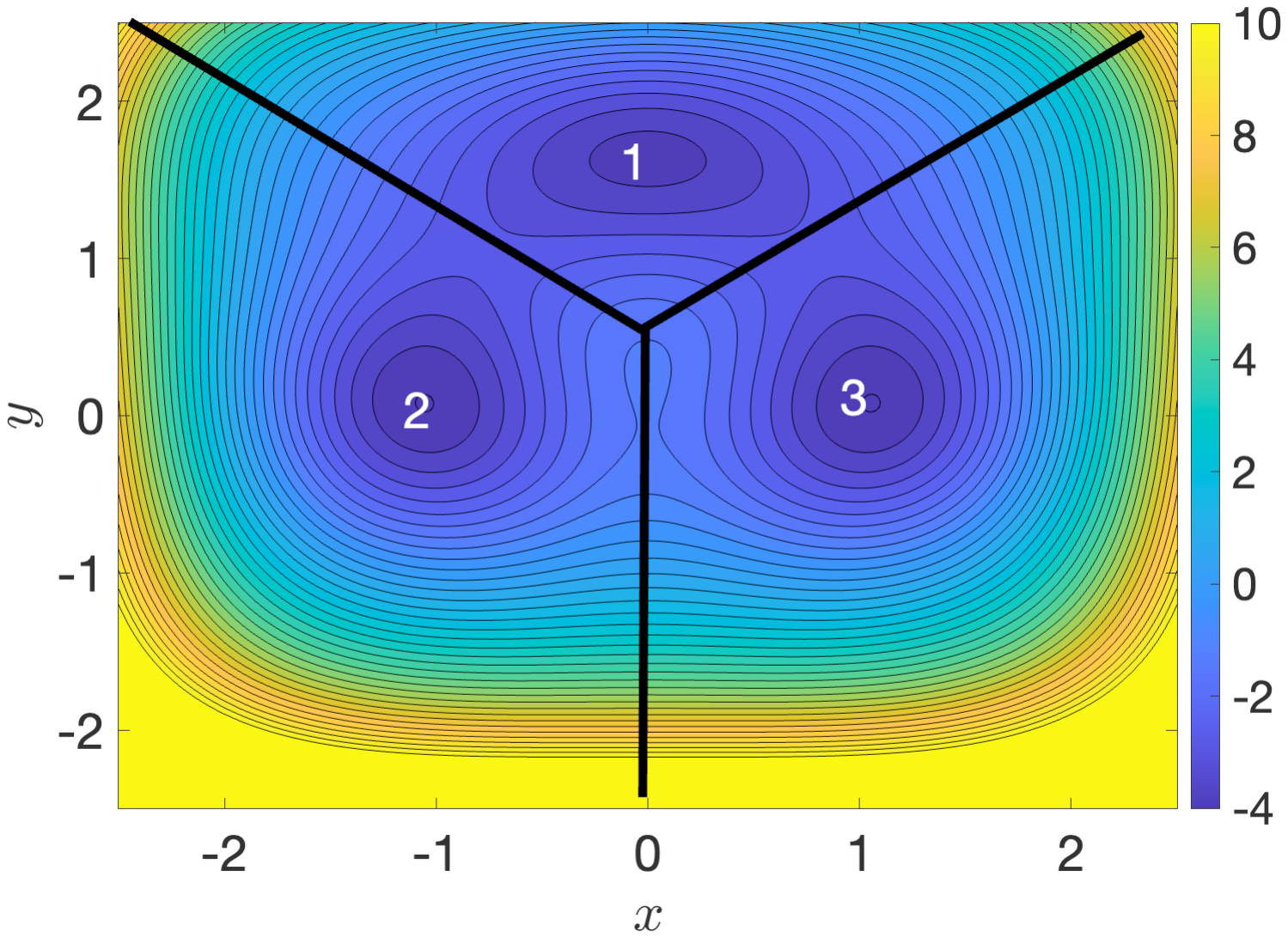}}    b)\subfigure{\includegraphics[width=0.47\textwidth]{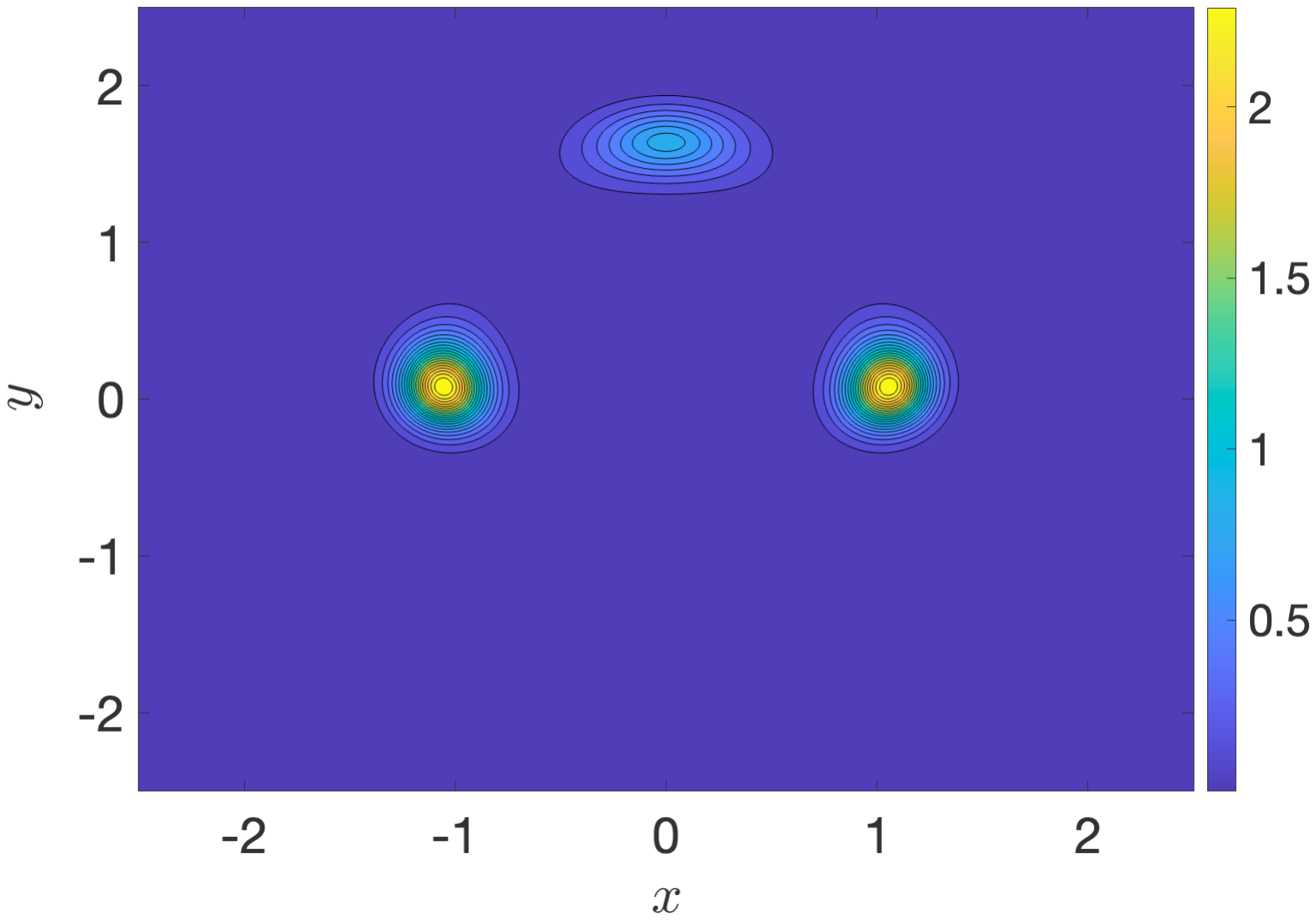}} \\
c)\subfigure{\includegraphics[width=0.47\textwidth]{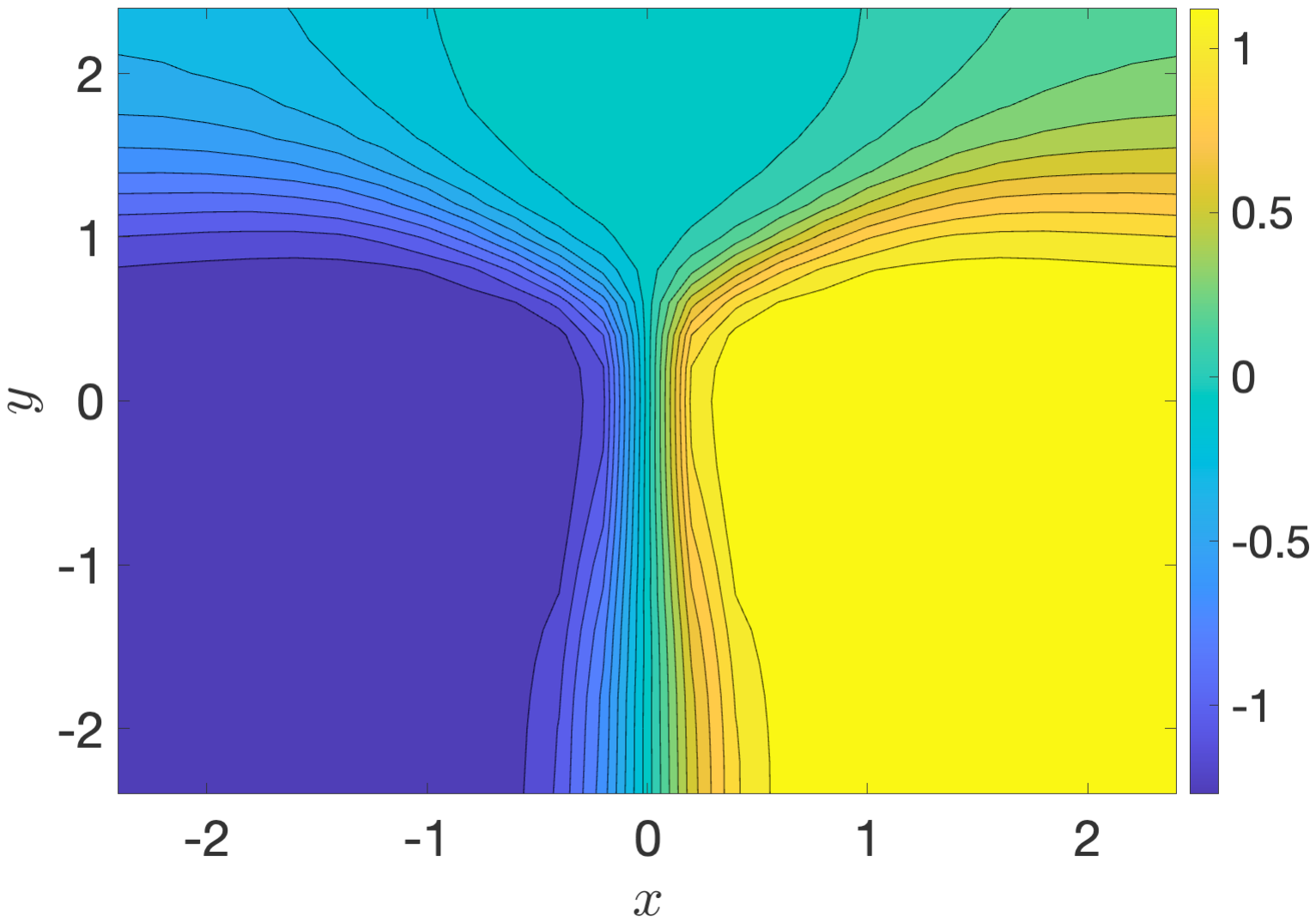}}    d)\subfigure{\includegraphics[width=0.47\textwidth]{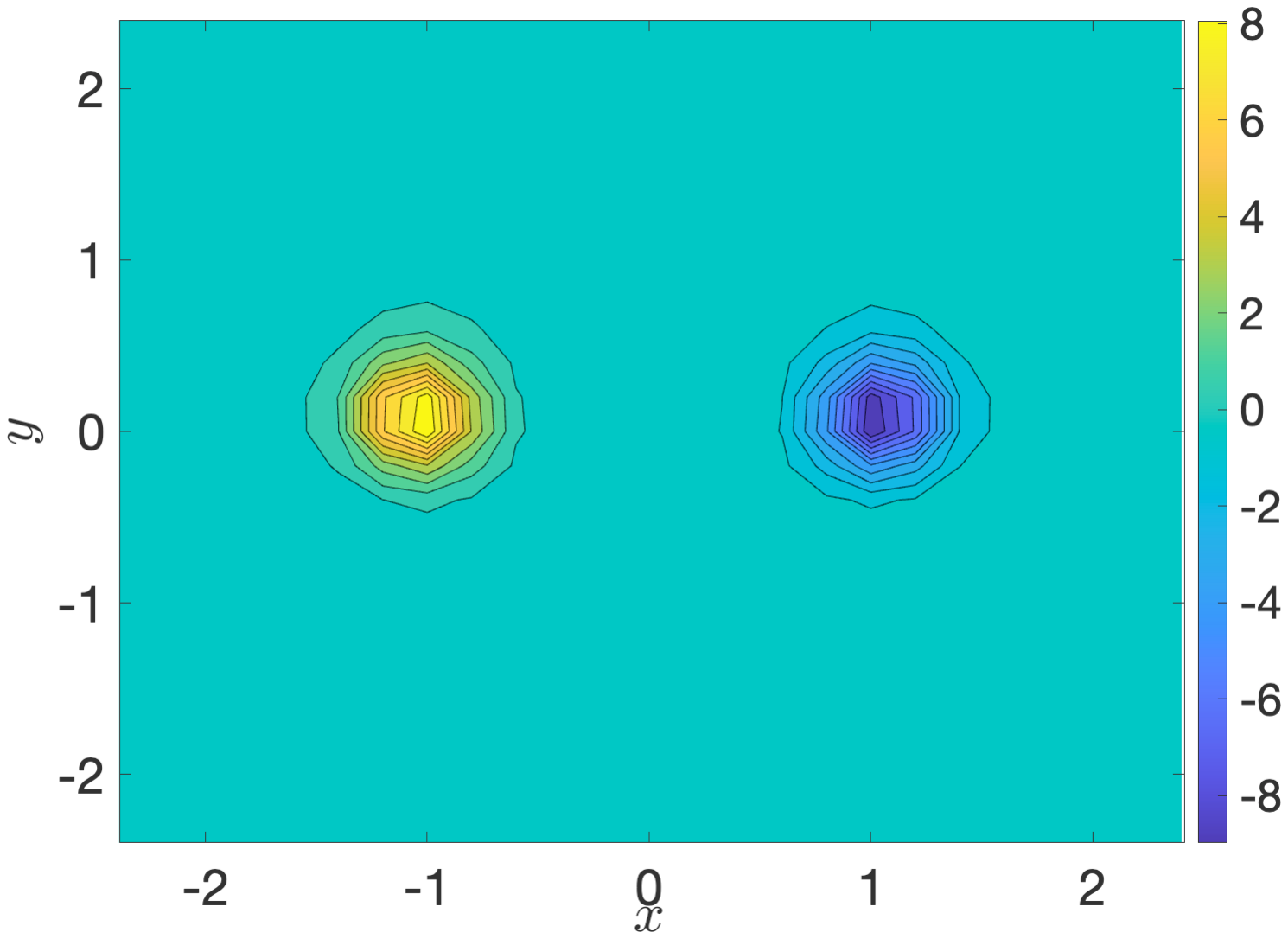}} \\
e)\subfigure{\includegraphics[width=0.47\textwidth]{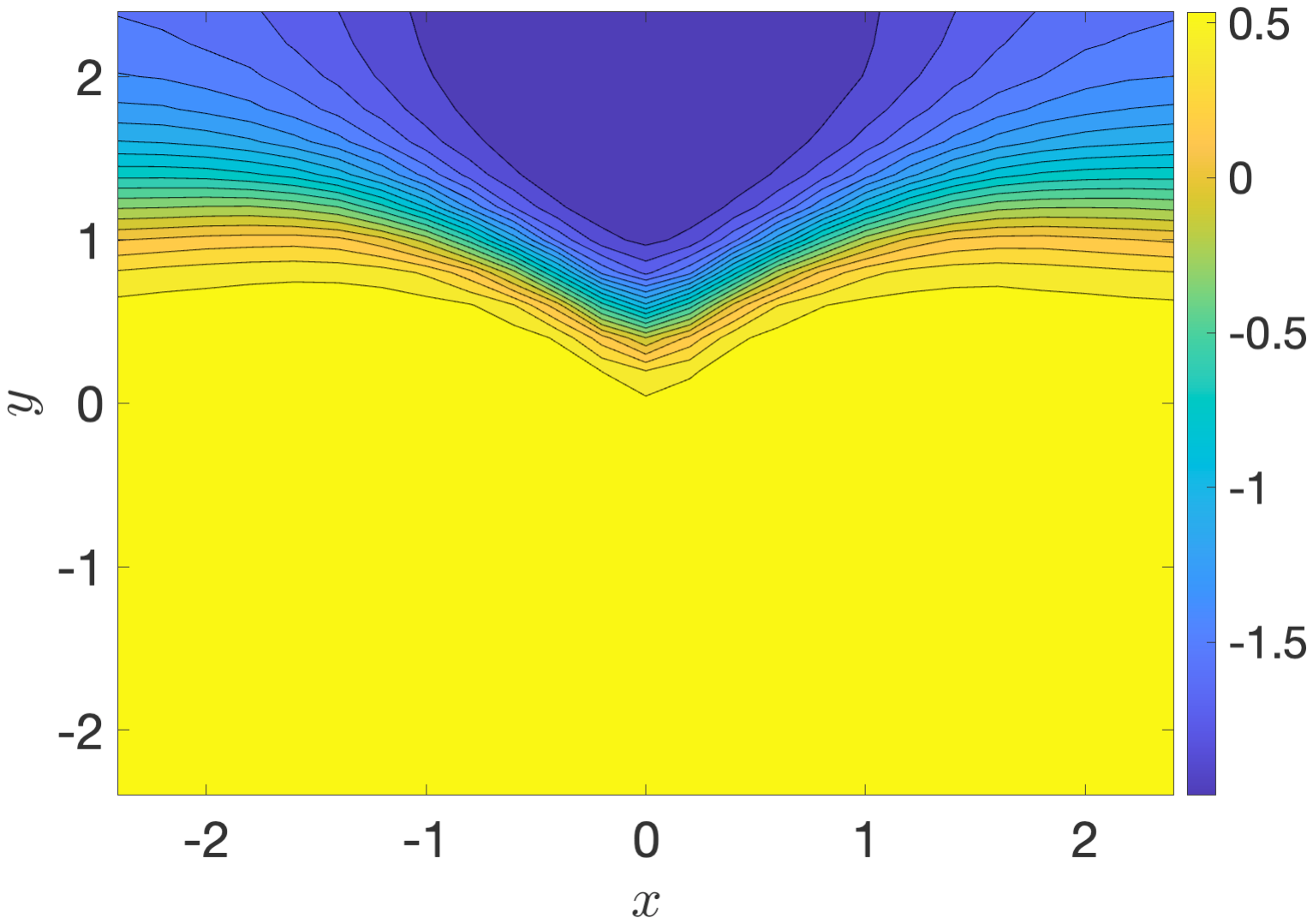}}    f)\subfigure{\includegraphics[width=0.47\textwidth]{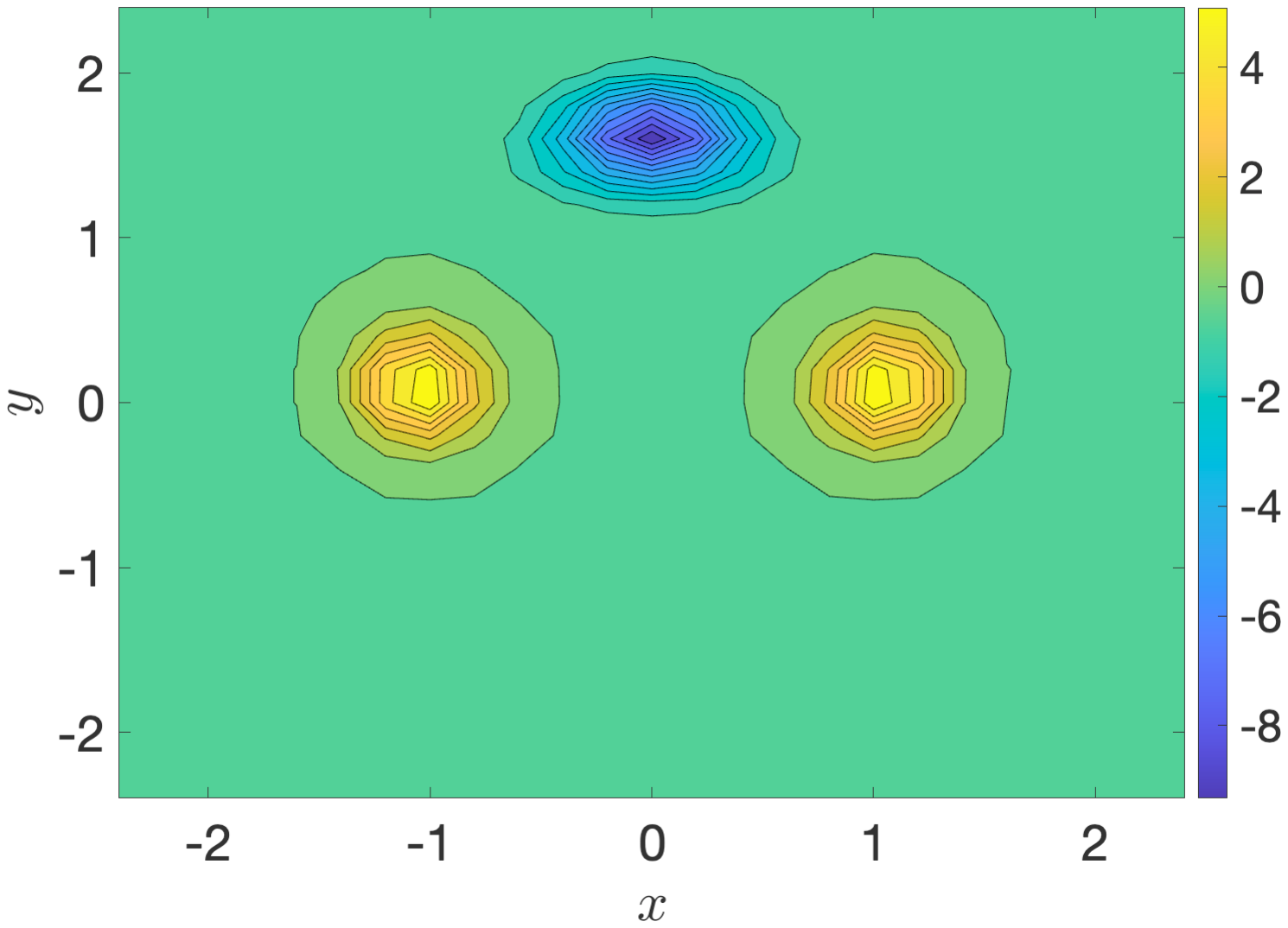}} \\
    \caption{(a) Potential function $V(x,y)$ with approximate indication of the three quasi-invariant regions surrounding the minima of the $V$.(b) Invariant Measure $\rho_0\propto\exp(-2V(x,y)/\sigma^2)$. (c). First subdominant Koopman mode, $\lambda_1=0.9916$.  (d) First subdominant mode of the Perron-Frobenius Operator. (e) Second subdominant Koopman model $\lambda_2=0.9655$.  (f) Second subdominant mode of the Perron-Frobenius Operator.  }
    \label{modes}
\end{figure}

We wish to provide here a proof of concept to test the usefulness of the framework proposed earlier. We consider the following 2-dimensional (2D) Langevin equation:
\begin{align}\label{Langevin}
    \mathrm{d}x=F_x(x,y)+\sigma \mathrm{d}W_1=-\partial_xV(x,y)+\sigma \mathrm{d}W_1\\
    \mathrm{d}y=F_y(x,y)+\sigma \mathrm{d}W_2=-\partial_yV(x,y)+\sigma \mathrm{d}W_2
\end{align}
where $\mathrm{d}W_1$ and $\mathrm{d}W_2$ are increments of independent Wiener processes, {\color{black}and the drift is defined as minus the gradient of the following potential:}  
\begin{align}
V(\mathbf{x})=V(x,y)&=3\exp(-x^2-(y-1/3)^2)-3\exp(-x^2-(y-5/3)^2)\nonumber\\&-5\exp(-(x+1)^2-y^2)-5\exp(-(x-1)^2-y^2)\nonumber\\
&+1/5x^4+1/5(y-1/3)^4-y.
\end{align}
{\color{black}Given the choice of the drift and of the noise law, we are considering here an equilibrium system obeying detailed balance \cite{pavliotisbook2014}.} The potential is depicted in Fig. \ref{modes}a and features three local minima, located at $(x_1,y_1)\approx (0,1.55)$,  $(x_2,y_2)=(-1,0)$, and $(x_3,y_3)=(1,0)$. The potential we consider here corresponds almost exactly to a case study presented in \cite{Klus2016}. We have included an additional term $-y$ in the definition of the potential in order to reduce the value of $V$ at the local minimum in $(x_1,y_1)$ to a value closer to the absolute minimum that is realized at $(x_2,y_2)$ and $(x_3,y_3)$.

The Fokker-Planck equation associated with the  previous Langevin equation reads as \cite{risken,pavliotisbook2014}:
\begin{equation}\label{FPlanck}
   \partial_t\rho(\mathbf{x},t) = \mathcal{L}_0 \rho(\mathbf{x},t) = \partial_x  (\partial_x V(\mathbf{x}) \rho(\mathbf{x},t))+\partial_y(\partial_y V(\mathbf{x}) \rho(\mathbf{x},t)) +\frac{\sigma^2}{2} (\partial_x^2+\partial_y^2)\rho(\mathbf{x},t)   
\end{equation}
{\color{black}The solution can be expressed as $\rho(\mathbf{x},t)=\exp(\mathcal{L}_0 t)\rho_{in}(\mathbf{x})$ where $\rho_{in}(\mathbf{x})$ is the initial condition and $\exp(\mathcal{L}_0 t)$ is the strongly continuous semigroup generated by $\mathcal{L}_0$, which is densely defined in a suitable Banach space \cite{pavliotisbook2014,Chekroun_al_RP2}. The Perron-Frobenius operator $P^{t,0} =\exp (\mathcal{L}_0t)$ describes the evolution of measures for a time $t$. Since we have a confining potential for the drift  and non-degenerate noise, we have a unique invariant measure that is absolutely continuous with respect to Lebesgue, so that it can be written as $\rho_0\mathrm{d}\mathbf{x}$, where $\mathcal{L}_0\rho_0=0$ \cite{pavliotisbook2014,Bolley2012}. We obtain $\rho_0(\mathbf{x})=Z^{-1} \exp(-2V(\mathbf{x})/\sigma^2)$, where $Z$ is the normalisation factor ensuring that $\int \mathrm{d}\mathbf{x}\rho_0(\mathbf{x})=1$. As a result, the natural functional space for our problem is given by $L^2_{\rho_0}$. The Koopman operator $K^{t,0}$ on measurable observables and is defined as the adjoint of the Perron-Frobenius operator: $\int K^{t,0} \Psi(\mathbf{x})   \rho_{in}(\mathbf{x})\stackrel{\text{def}}{=}\int \mathrm{d}\mathbf{x}\Psi(\mathbf{x}) P^{t,0} \rho_{in}(\mathbf{x})$, where $K^{t,0}=(P^{t,0})^T=\exp(\mathcal{L}^T_0 t)$ and $\mathcal{L}^T_0 \Psi(\mathbf{x}) = - \partial_x V(\mathbf{x})  \partial_x \Psi(\mathbf{x})   - \partial_y V(\mathbf{x})  \partial_y \Psi(\mathbf{x})  + \frac{\sigma^2}{2} (\partial^2_x +\partial^2_y) \Psi(\mathbf{x})$.}

In what follows, we assume $\sigma=0.7$. Our numerical simulations are performed using the standard Euler-Maruyama scheme \cite{kloeden_1992} with $dt=0.05$. We sample our output every 20 times steps, so that our reference time scale is $\tau=1$. The dynamics given by Eqs. \ref{Langevin}-\ref{FPlanck} describes an equilibrium system.

To approximate \( P^{\tau,0}\), we discretize the phase space of the system into \( N \) disjoint subsets  \( \{B_1, B_2, \dots, B_N\} \), forming a partition of the space. The {Ulam transfer operator} \( P^{\tau,0}_{\{N\},ij} \), a finite-dimensional stochastic matrix describes the probability that the orbit of the system is at time $t+\tau$ in the subset $B_i$ is at time $t$ it was in the subset $B_j$. In our case, the $B_j$'s are given by the 625 squares with side 0.2 centred around the origin. Whilst in principle one would need to cover the entire $\mathbb{R}^2$ the confining potential makes all regions beyond those we consider here entirely irrelevant unless one consider extremely long time scales. 

Following \cite{Klus2016}, we populate each cube with 1000 ensemble members distributed uniformly according to the Lebesgue measure. Each member evolves for $\tau=1$ time unit. We then construct an estimate of $P^{\tau,0}_{\{N\}}$ by counting the transitions. We repeat the operation 20 times and by averaging we obtain our best estimate of $P^{\tau,0}_{\{N\}}$, which constitutes our reference discretized stochastic matrix, so that  $\mathcal{M}= P^{\tau,0}_{\{N\}}$. The solution to $\mathcal{M}\nu_{inv}=\nu_{inv}$ gives an accurate  approximation  in the gridded domain defined by the $B_j$'s of the true invariant measure $\rho_0$. 

It is possible to filter out the  dynamics associated with spurious non-equilibrium modes due to the  finite sampling by proceeding as follows. First, we compute the matrix $\mathcal{N}_{ij}=\mathcal{M}_{ji}\frac{\nu_{inv,i}}{\nu_{inv,j}}$. Them we update our estimate of  $\mathcal{M}$ as follows: ${\mathcal{M}}\rightarrow\frac{\mathcal{M}+\mathcal{N}}{2}$, so that detailed balance is enforced in the finite-state representation of the system\footnote{{\color{black}If $\mathcal{M}$ obeys detailed balance, the symmetric matrix $h=D \mathcal{M} D^{-1}$, where $D=diag(\nu_{inv,1},\ldots,\nu_{inv,N})$ has the same eigenvalues as those of $\mathcal{M}$. Estimating the eigenvalues via $h$ could prove helpful for large $N$ thanks to its numerically convenient symmetry property.}}.

Whilst the first eigenvector of $\mathcal{M}$ is shown in Fig. \ref{modes}b, the first right eigenvector of the Koopman operator $\mathcal{M}^T$, is constant everywhere. We obtain 624 additional eigenvalue-eigenvector pairs for the matrix $\mathcal{M}$. 
The two subdominant eigenvectors are depicted in Fig. \ref{modes}d and Fig. \ref{modes}f. The first one describes the transitions between the neighbourhoods of $(x_2,y_2)$ and $(x_3,y_3)$. The second one describes the transitions between the neighbourhoods of $(x_1,y_1)$ and either $(x_2,y_2)$ or $(x_3,y_3)$. The corresponding subdominant eigenvectors of the Koopman operator are depicted in Fig. \ref{modes}c and Fig. \ref{modes}e. There is a very large spectral gap between the three dominant modes (corresponding to $\lambda=1$, $\lambda\approx0.9917(1)$, $\lambda\approx0.9655(1)$ the rest of the spectral components ($\lambda_4\approx0.050(1))$, which indicates that the system can safely undergo a model reduction procedure. This could be algebraically achieved by substituting $\mathcal{K}\rightarrow \sum_{i=1}^3\lambda_i\Pi_i$, and $\mathcal{M}\rightarrow \sum_{i=1}^3\lambda_i\Pi_i^T$. Additionally, the level sets of the two subdominant Koopman modes can be used to identify the three metastable regimes of the system, which are indicated as (1), (2) and (3) in Fig. \ref{modes}a and correspond to the basins of attraction of the three local minima in the case the stochastic forcing is switched off. Performing k-means clustering \cite{Forgy65,Lloyd82} of the data choosing 3 as number of clusters leads to the same partition of the data space, whereby the three regions (1), (2), (3) are the  Vorono\"i cells \cite{Aurenhammer1991} associated with the cluster centers.

We  use such a geometrical partition of the phase space to introduce a separate reduced-order  representation of the dynamics, whereby the phase space of the system is partitioned into  three states, corresponding to the regions (1), (2), and (3). In the simple case described here, this corresponds to MSM. This amounts to neglecting entirely the intrawell dynamics, which, in the Ulam description above, is mostly captured by the Koopman modes with index $\geq4$. We estimate the reduced Markov model (RMM) discretized transfer operator ${M}$ by performing a single run lasting $10^7$ time units (after disregarding a small transient) and counting the transitions between the 3 states described above. {\color{black}We then enforce detailed balance as above.} The obtained estimates for ${M}$ and its eigenvectors are given below, where all the numbers indicated below have an approximate uncertainty of 1 in the last digit we have written out.

\[
\tilde{M}=\left( \begin{array}{ccc}
0.9701 & 0.0095 & 0.0095 \\
0.0149 & 0.9904 & 0.00006 \\
0.0149 & 0.00006 & 0.9904
\end{array} \right)
\quad \nu_{inv} = \begin{pmatrix} 0.240 \\ 0.380 \\ 0.380 \end{pmatrix} 
\quad \nu_2 = \begin{pmatrix} 0.000 \\ 0.701 \\ -0.701 \end{pmatrix} 
\quad \nu_3 = \begin{pmatrix} -0.810 \\ 0.405 \\ 0.405 \end{pmatrix}
\]
The two nontrivial eigenvalues $\lambda_2=0.9904(1)$ and $\lambda_3=0.9606(1)$ correspond closely to the first two subdominant eigenvalues obtained using the Ulam method, and the corresponding eigenvectors provide a coarse grained version of the figures provided in Fig. \ref{modes}d and Fig. \ref{modes}f.

\subsection{Applying Extra Forcings}
We now consider the following perturbation to the drift term: $(F_x(x,y),F_y(x,y))\rightarrow( F_x(x,y) +\epsilon_1 f_1(t), F_x(x,y +\epsilon_1 f_2(t))$, where $f_1(t)$ and $f_2(t)$ give the  time modulation of the forcing with $|f_1(t)|, |f_2(t)|\leq1$ and $\epsilon_1,\epsilon_2$ are the bookkeeping parameters controlling the intensity of the applied perturbation. We set ourselves in the regime of linear response, so that it is natural to assume $|\epsilon_1|,|\epsilon_2|\ll1$. The time-dependent expectation value of a general observable $\Psi(x,y)$ can be written as
\begin{equation}
    \langle \Psi\rangle(t)=\langle \Psi\rangle_0+\epsilon_1 \int_{-\infty}^\infty \mathrm{d}t_1 G^{(1)}_{x,\Psi}(t-t_1)f_1(t_1)+\epsilon_2 \int_{-\infty}^\infty \mathrm{d}t_1 G^{(1)}_{y,\Psi}(t-t_1)f_2(t_1)+h.o.t.
\end{equation}
where $\langle \Psi\rangle_0=\int \mathrm{d}x\mathrm{d}y\rho_0(x,y)\Psi(x,y)$ is the expectation value of $\Psi$ in the unperturbed  state, whilst 
\begin{equation}\label{FDT}
    G^{(1)}_{x/y,\Psi}=-\Theta(t)\langle \partial_{x/y}\log(\rho_0)\Psi(t)\rangle_0=\frac{2}{\sigma^2}\Theta(t)\langle \partial_{x/y}V \Psi(t)\rangle_0
\end{equation}
are the causal Green's functions for the observable $\Psi$ associated with the perturbations acting along the $x$ and $y$ directions, respectively. These formulas can be readily derived from the general version of the FDT \cite{marconi2008fluctuation,Santos2022}. 

It is possible to associate the applied perturbation to the vector field to changes in the discretized Perron-Frobenius operators constructed according to the protocols above. We define $m_x$ ($m_y$) the perturbation matrix associated with the extra push in the $x$ ($y$) direction, so that $\mathcal{M}\rightarrow\mathcal{M}+\epsilon_1m_xf_1(n)+\epsilon_2m_yf_2(n)$. 

\subsubsection{Linear Response - Observables}
In order to estimate the matrices $m_x$ for the Ulam discretization, we repeat  the same protocol considering the perturbed dynamics realised by choosing $\epsilon_1=0.05$ and $f=1$. We derive $P^{\tau,+}_{\{N\}}$. We repeat the experiment by choosing $\epsilon_1=-0.05$ and $f=1$, and derive $P^{\tau,-}_{\{N\}}$. We estimate $m_x=(P^{\tau,+}_{\{N\}}-P^{\tau,-}_{\{N\}})/(2\epsilon_1)$. Note that using centred differences ensures high precisions when studying linear response \cite{gritsun2017}. In order to estimate $m_y$, we repeat the same procedure described above by considering  $\epsilon_2=-0.05$ and $\epsilon_2=-0.05$, respectively. 

Similarly, we estimate $m_x$ and $m_y$ for 3-state Markov model by performing long simulations of duration $10^7$ time units with perturbed dynamics, by estimating the Perron-Frobenius operator in each case, and by taking the centred differences. We use the same value $|\epsilon_1|=|\epsilon_2|=0.05$. We have verified in all cases that this is accurately within the regime of linearity of the system's response. 

We choose as observables $\Psi_1=x$ and $\Psi_2=y$. We first note that because of the symmetry of the system $G^{(1)}_{x,y}(\tau)=G^{(1)}_{y,x}(\tau)=0$. We then focus on the case where we force the system in the $j$ direction and use $j$ as observable, $j=x,y$. We then estimate  $G^{(1)}_{x/y,x/y}$ through formula \ref{FDT} by collecting statistics for 500 independent ensemble runs of the unpertubed system each lasting $10^6$ time units. Note that this 50 times as many data as those used for constructing the Markov chains. From the knowledge of $m_x,$, $m_y$, it is straightforward to compute the Green's functions for $\Psi_1=x$ and $\Psi_2=y$ using Eq. \ref{Greenm}. 

\begin{figure}
    \centering
\includegraphics[width=0.8\textwidth]{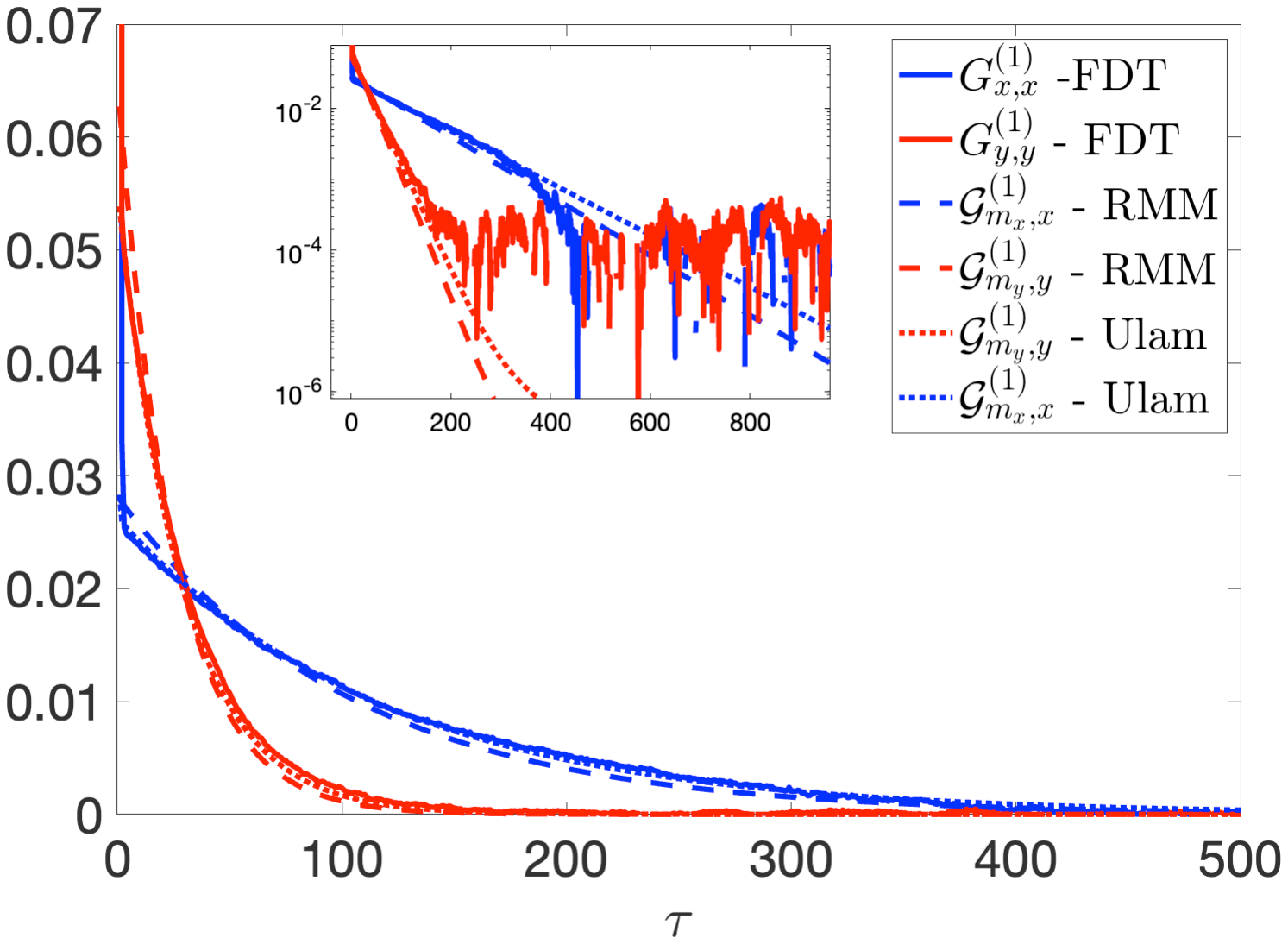}\\
    \caption{Linear Green's function for the $x$ and $y$ observables for additive forcing acting on $x$ (index $m_x$) or $y$ (index $m_y$) direction. Results are shown for the FDT estimate and the estimates obtained using Markov models constructed with Ulam's discretization and the 3-state RMM. The inset emphasizes the exponential decay of the Green's functions.}
    \label{Greensfigure}
\end{figure}

The estimates we obtain for the Green's functions of interest are shown in Fig. \ref{Greensfigure}. When considering RMM, we clearly see that $\mathcal{G}^{(1)}_{m_x,x}(\tau)\propto \lambda_2^\tau$ and $\mathcal{G}^{(1)}_{m_y,y}(\tau)\propto\lambda_3^\tau$: as a result of the choice of forcing/observable pair, only one Koopman mode is retained in the expansion given in Eqs. \ref{Gdecompo}- \ref{KoopmanResponseb}. Indeed, it is clear that if we force along the $x$-direction and choose $x$ as observable, the second Koopman mode (or only the second Perron-Frobenium mode) is the only one that given a nonnvanishing contribution. Similarly, the third mode is the only one retained in the spectral expansion of the Green's function when forcing along  $y$-direction and choosing $y$ as observable. 

Remarkably, also when considering the much higher complexity Markov chain constructed through the Ulam method, to a very good approximation we have $\mathcal{G}^{(1)}_{m_x,x}(\tau)\propto \lambda_2^\tau$ and $\mathcal{G}^{(1)}_{m_y,y}(\tau)\propto\lambda_3^\tau$ (note that the corresponding $\lambda$'s are slightly larger than in the RMM case). This means that by and large only one of the 625 natural modes of variability of the system matters in defining the response to perturbation, at least in the cases we consider here. This shows that the Koopman decomposition provides the much desired property of interpretability of the response. 
It is apparent that the estimates of the two Green's functions obtained using the FDT are relatively noisy, despite the use of a much larger dataset than what has been used for constructing the Gree's function via Markov chains. A side remark is that, by construction, $\lim_{t\rightarrow0^+}\mathcal{G}^{(1)}_{m_x,x}(t)=\lim_{t\rightarrow0^+}\mathcal{G}^{(1)}_{m_y,y}(t)=1$. Nonetheless, the function collapses to much lower values within one time unit because of the very rapid decay of correlation due to the rapidly decaying Kolmogorov modes of the continuum system. As soon as we consider $t\geq1$, a rather good agreement is found with the estimates obtained via Markov chains. 

\subsubsection{Linear Response - Correlations}

\begin{figure}
    \centering  a)\subfigure{\includegraphics[width=0.47\textwidth]{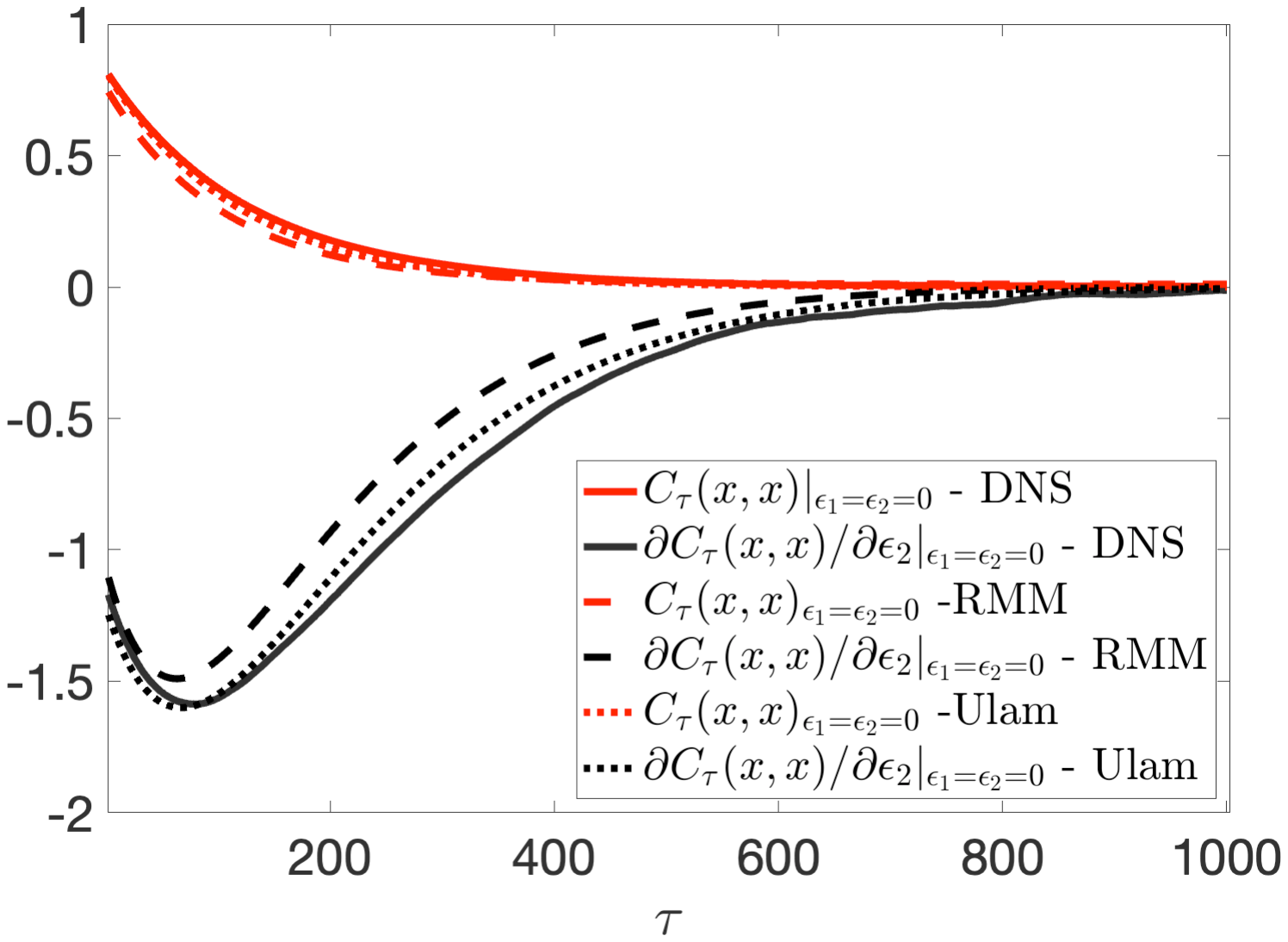}}    b)\subfigure{\includegraphics[width=0.47\textwidth]{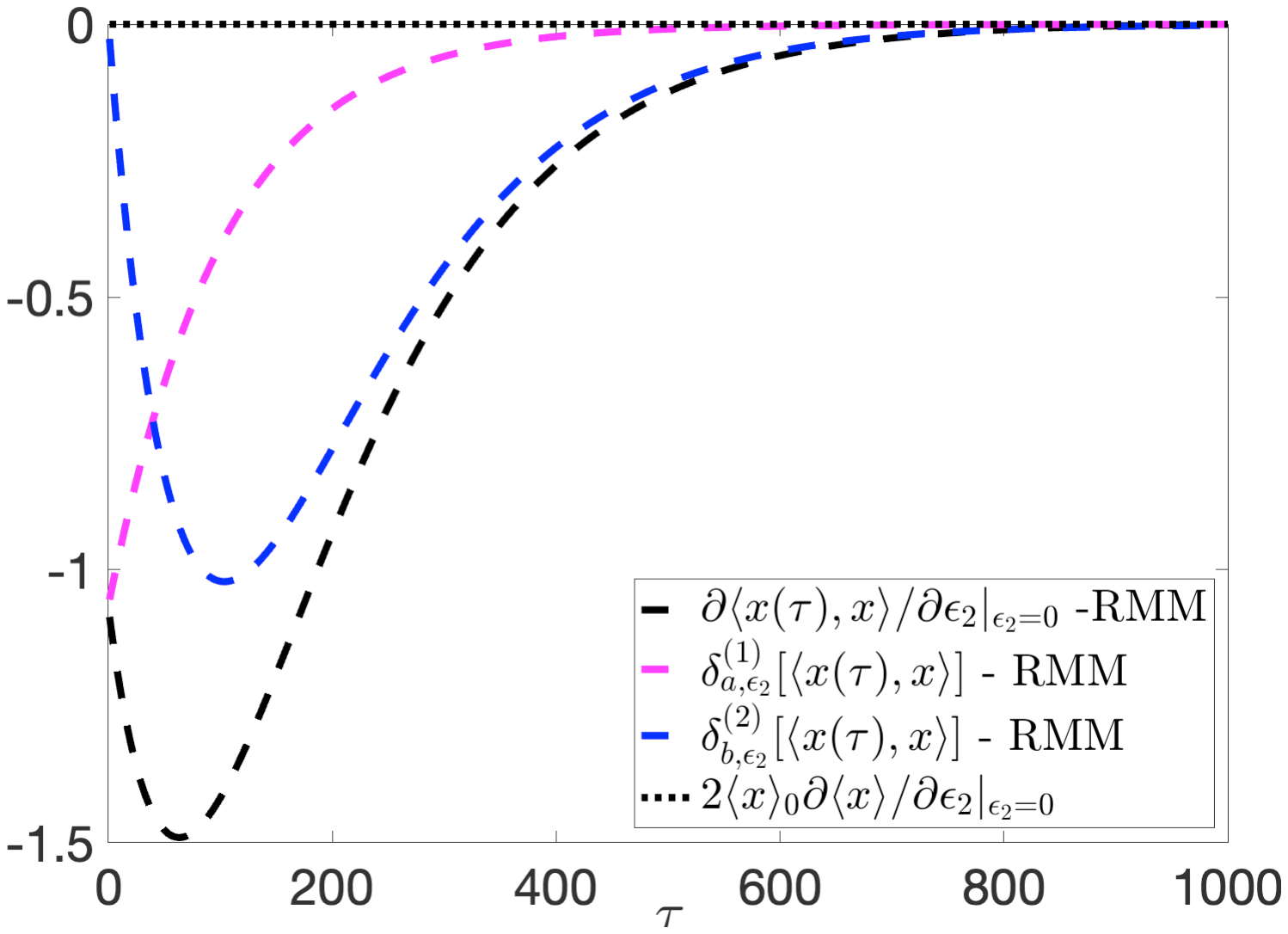}} \\
c)\subfigure{\includegraphics[width=0.47\textwidth]{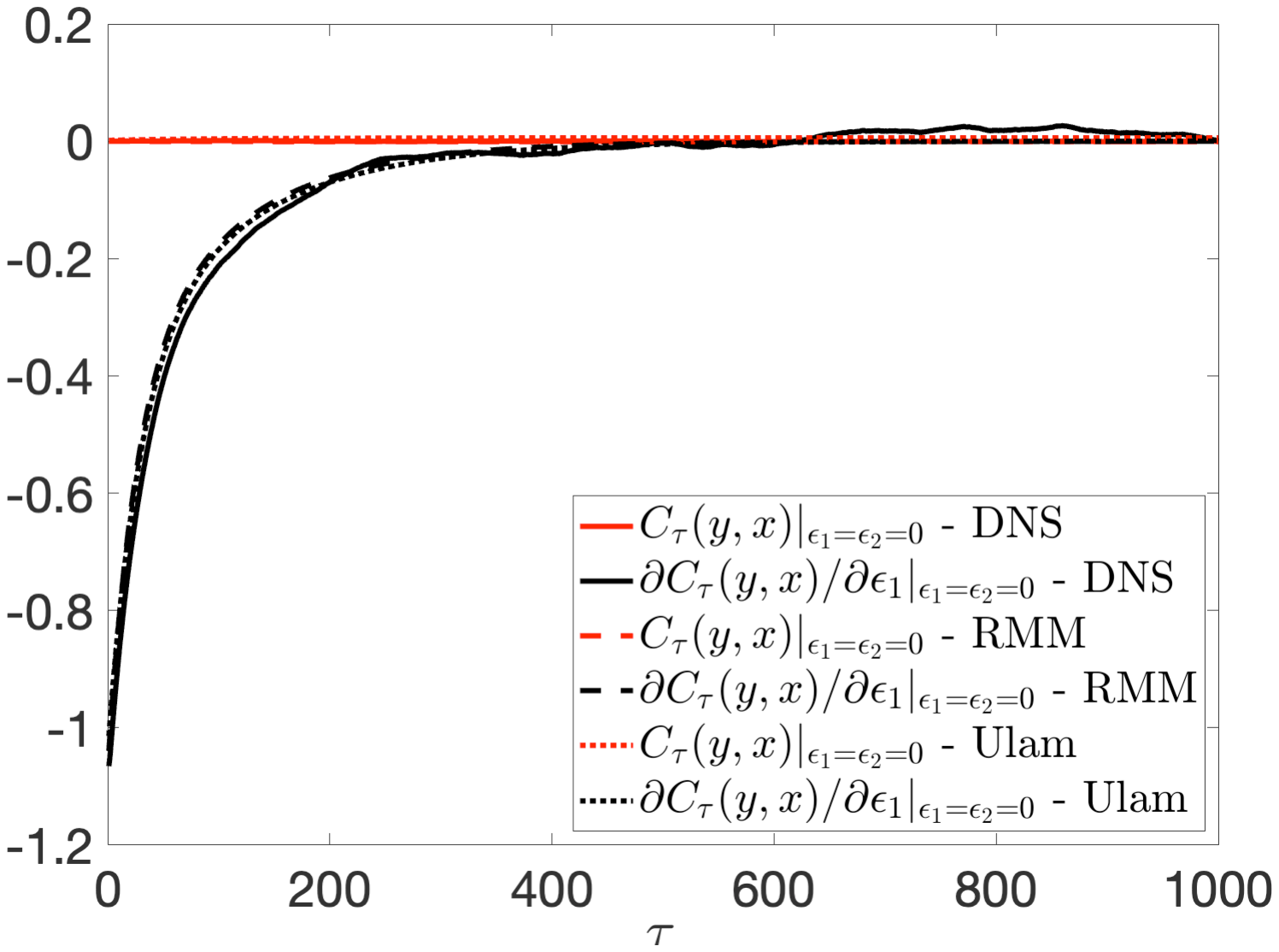}} 
d)\subfigure{\includegraphics[width=0.47\textwidth]{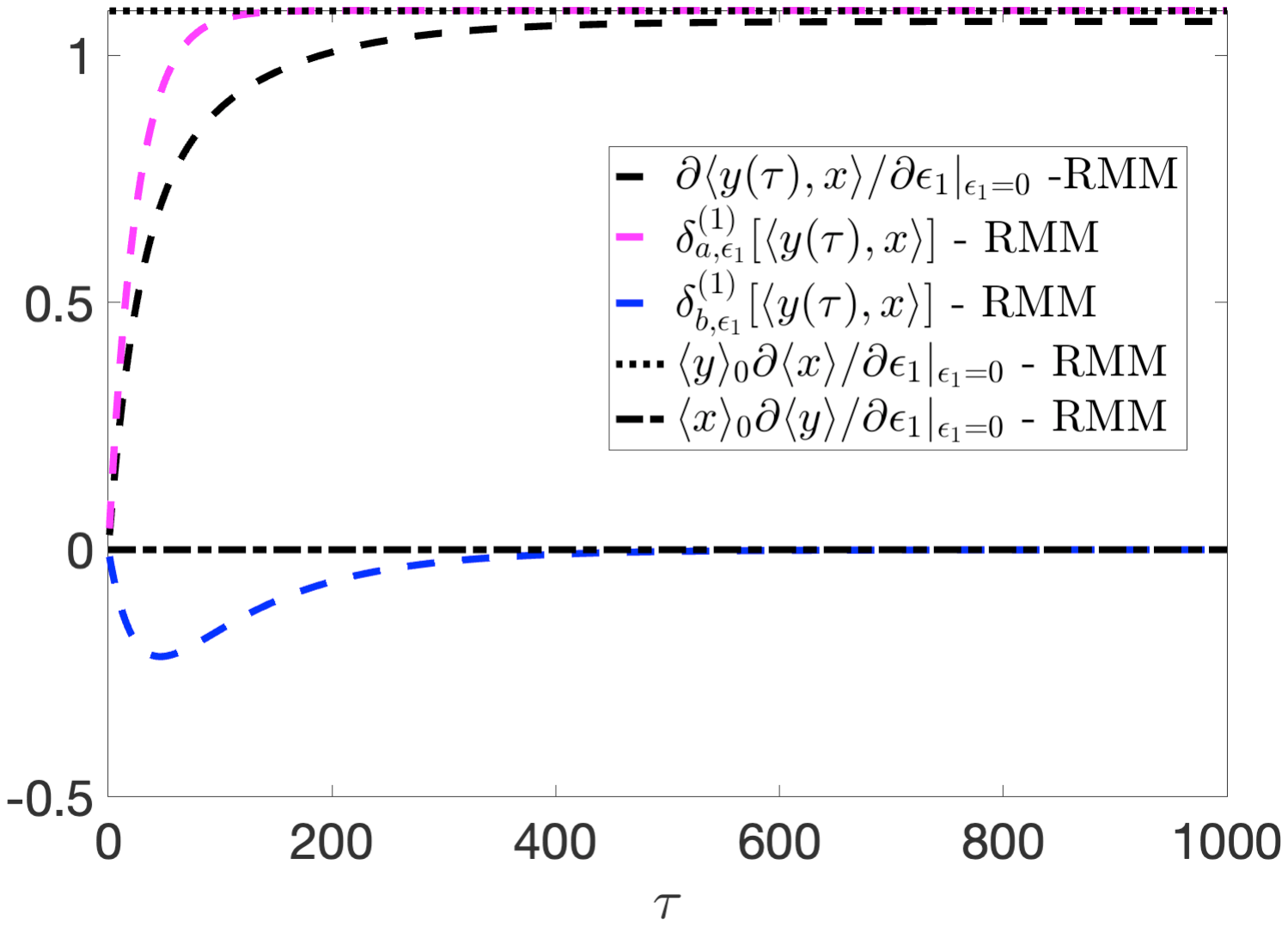}} \\
e)\subfigure{\includegraphics[width=0.47\textwidth]{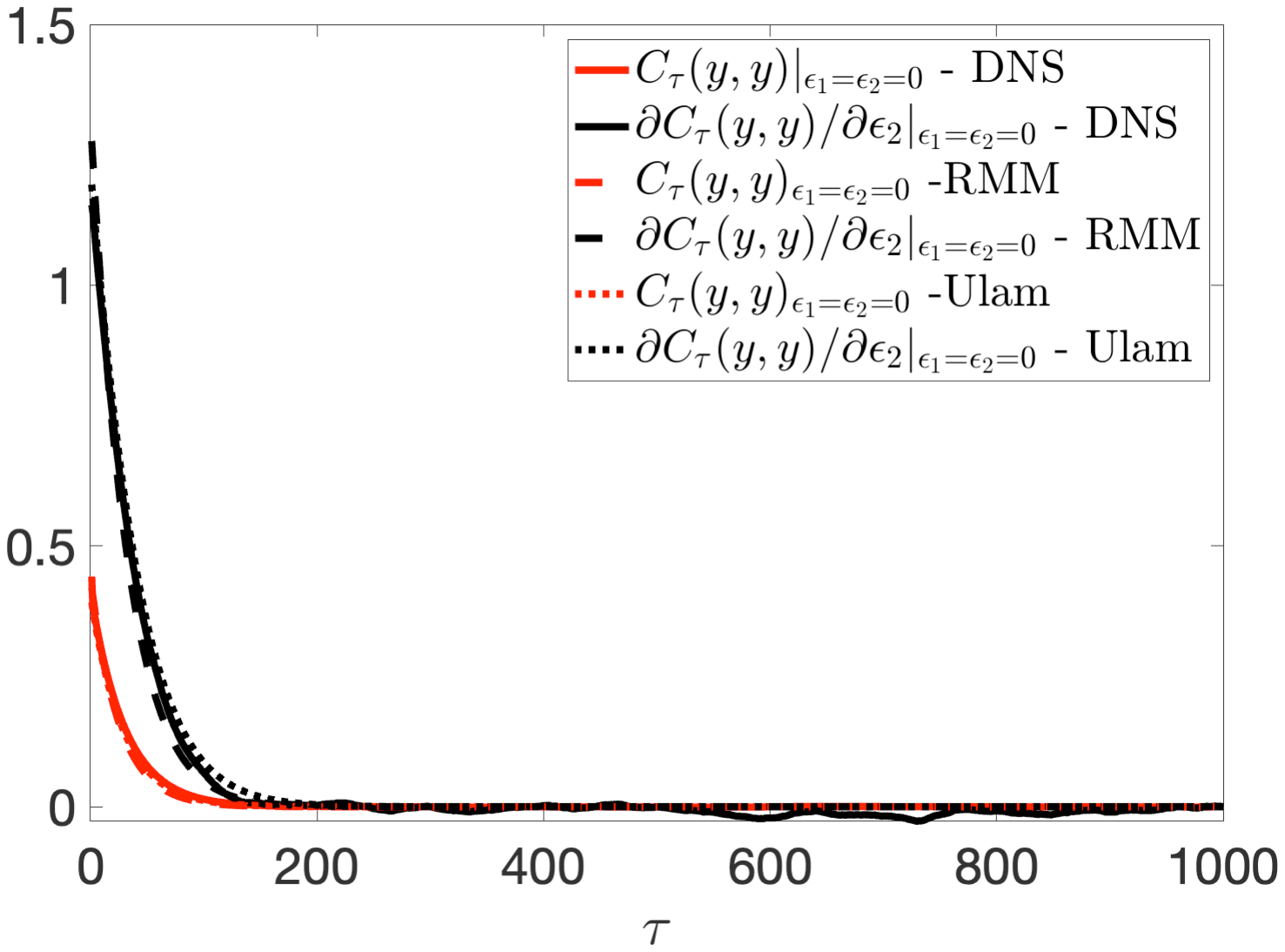}} 
f)\subfigure{\includegraphics[width=0.47\textwidth]{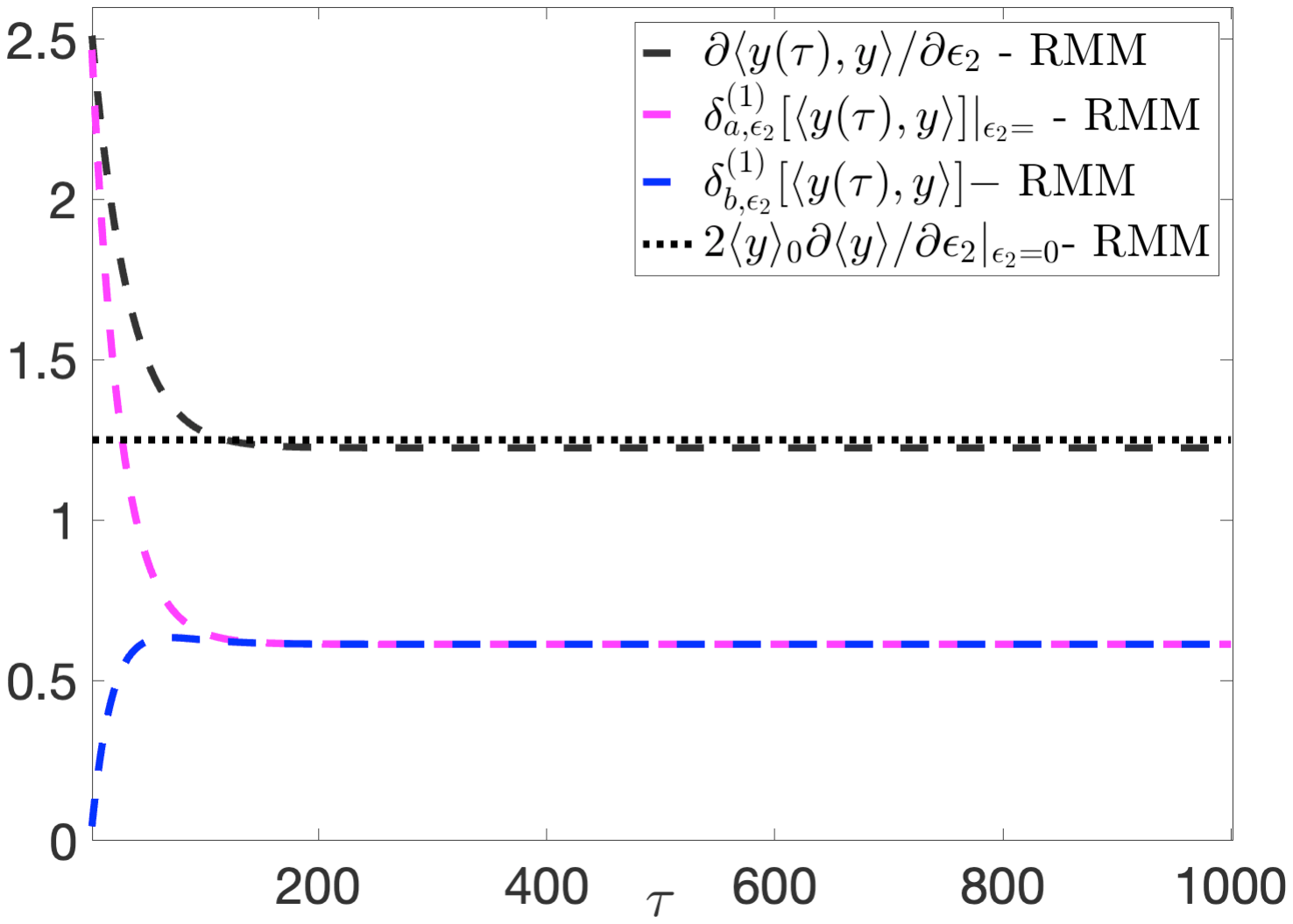}}
    \caption{(a) Estimate of $C_\tau(x,x)\rangle$ via direct numerical simulations (DNS), RMM, and Ulam method. Red lines: Reference state. Black lines: sensitivity with respect to $\epsilon_2$. (b) Decomposition of the linear response in (a) in the four terms discussed in Eq. \ref{decomporesponse}  (RMM). (c) Same as (a), but for $C_\tau(y,x)$ and its sensitivity with respect to $\epsilon_1$. (d) Same as (b), in reference to the linear response shown in (c). (e) Same as (a), but for $C_\tau(y,y)$ and its sensitivity with respect to $\epsilon_2$. (f). Same as (b), in reference to the linear response shown in (e). }
    \label{correlations}
\end{figure}

Next we venture into the analysis of correlations and of their response to perturbations for the full system and for its discretized representation via Markov chain. We consider the time-lagged correlations $C_\tau(x,x)$, $C_\tau(y,x)$, and $C_\tau(y,y)$. Our results are shown in Figs. \ref{correlations}a)-f). Note that since the potential $V$ of the unperturbed system and the observable x have opposite symmetry with respect to the exchange $x\rightarrow-x$, we derive that $C_\tau(y,x)|_{\epsilon_2=\epsilon_1=0}=0$, $\partial C_\tau(y,x)/\partial \epsilon_2|_{\epsilon_2=\epsilon_1=0}=0$,  $\partial C_\tau(x,x)/\partial \epsilon_1|_{\epsilon_2=\epsilon_1=0}=0$, and $\partial C_\tau(y,y)/\partial \epsilon_1|_{\epsilon_2=\epsilon_1=0}=0$.

In the case of the full system, we adopt a simple strategy of direct numerical simulation (DNS). The correlations have been computed by collecting statistics along the same simulation used to estimate the Green's functions above. In order to evaluate their response to perturbations, we have run two additional simulations with $f_1=1$, $f_2=0$ and $\epsilon_1=\pm0.05$ plus two additional simulations with $f_1=0$, $f_2-=1$ and $\epsilon_2=\pm0.05$ and taken centred differences  to estimate the linear response to perturbation in the $x$ and $y$ directions.  

In the case of the Markov chain, we have used the expression of correlations and the linear response formulas to static forcings provided in Sect. \ref{responsecorrelations}. 

We obtain that - see Figs. \ref{correlations}a), \ref{correlations}a)c), and \ref{correlations}e) - there is a very good agreement between the correlations computed on the full system and those estimated via Markov chains. Comparing with Fig. \ref{Greensfigure}, it is apparent that $C_\tau(x,x)|_{\epsilon_2=\epsilon_1=0}\propto\exp(\lambda_2t)$ and $C_\tau(y,y)|_{\epsilon_2=\epsilon_1=0}\propto\exp(\lambda_3t)$.

Similarly, a good agreement is found when considering the sensitivities with respect to $\epsilon_1$ and $\epsilon_2$. In all cases it is apparent that a little fraction of the signal is lost when performing the coarse graining. The good performance obtained even in the case of the 3-state system indicates the effectiveness of the reduced order modelling strategy. Using the decomposition presented in Eq. \ref{decomporesponse}, we are able to separate the change in the correlation function in four components, which are associated with fundamentally different dynamical processes. Two terms come from the sensitivity in the expectation value of the  two observables we are considering. An additional term - indicated by $\delta^{(1)}_{b,\epsilon}$ comes from the change in the expectation value of the lagged product of the observable due exclusively to the change in  the measure (where instead the evolution occurs according to the unperturbed dynamics). The most interesting term is undoubtedly $\delta^{(1)}_{a,\epsilon}$, which measures the impact of the change in the dynamics occurring up to the considered time lag. Indeed, this terms vanishes as $\tau\rightarrow0$ and is, as already observed in \cite{LucariniWouters2017}, a specific element of response formulas for correlations. The interplay between the two terms $\delta^{(1)}_{a,\epsilon}$ and $\delta^{(1)}_{b,\epsilon}$ is nontrivial. 

\subsubsection{Nonlinear Response}
Whilst explicit formulas for nonlinear Green's functions exist \cite{ruelle_nonequilibrium_1998,lucarini2008}, their numerical implementation is extremely challenging because of the convoluted structure of differential operators acting at different times. The nonlinear Green functions can be formally seen as Volterra kernels \cite{Orcioni2014} and can in principle be constructed using neural networks \cite{Wray1994}. 

Using the formalism developed here, we derive easily implementable and easily interpretable formulas for the second (see Subsect. \ref{response}\ref{second}) and well as the arbitrary order nonlinear response (see App. \ref{arbitrary}), whereby the way different intrinsic time scales of the system and the corresponding modes interact with each other and with the forcing is extremely clear. Hence, as a final step proof-of-concept analysis of our system we have computed the second order Green's functions for our system.

{\color{black}Indeed, we consider here a slightly more complicated setting than what has been earlier presented in Subsect. \ref{response}\ref{second}.} We consider the case where both forcings described above are applied. It is easy to derive that the second-order response can be written as:
\begin{align}\label{secondorderresponse}
    \delta^{(2)}\Psi(n)=\sum_{i,j=x,y}\sum_{k=-\infty}^\infty \sum_{p=-\infty}^\infty&\epsilon_i\epsilon_j\Theta(k)\Theta(p)  \langle m_i^T(\mathcal{M}^T)^p m_j^T(\mathcal{M}^T)^k \Psi ,\nu_{inv}\rangle \nonumber \\ & \times f_i(n-k-p-2)f_j(n-k-1)+h.o.t. 
\end{align}
where we can define the following Green's functions 
\begin{equation}\label{Green2mxmy}
\mathcal{G}^{(2)}_{m_i,m_j,\Psi}=\Theta(k)\Theta(p)  \langle m_i^T(\mathcal{M}^T)^p m_j^T(\mathcal{M}^T)^k \Psi ,\nu_{inv}\rangle
\end{equation}
describes the combined effect of first applying the perturbation described by $m_j$ and then of the perturbation described by $m_i$. One should note that in general $\mathcal{G}^{(2)}_{m_i,m_j,\Psi} \neq \mathcal{G}^{(2)}_{m_j,m_i,\Psi}$ if $m_j\neq m_i$, because the time ordering matters. In our case, the second order response depends in general on all of these four terms. 

Given the symmetry properties of the system, if we choose $x$ as observable $\mathcal{G}^{(2)}_{m_x,m_x,x}=\mathcal{G}^{(2)}_{m_y,m_y,x}=0$, whilst $\mathcal{G}^{(2)}_{m_x,m_y,x} $ and $\mathcal{G}^{(2)}_{m_y,m_x,x}$ are in general non-vanishing. This implies that if we do not apply forcings in both direction, the second order response of $x$ vanishes. Instead, if we choose $y$ as observable, we have $\mathcal{G}^{(2)}_{m_x,m_y,y}=\mathcal{G}^{(2)}_{m_y,m_x,y}=0$, whilst $\mathcal{G}^{(2)}_{m_x,m_x,y}$ and $\mathcal{G}^{(2)}_{m_y,m_y,y}=0$ are non-vanishing. 

\begin{figure}
    \centering
\includegraphics[width=0.9\textwidth]{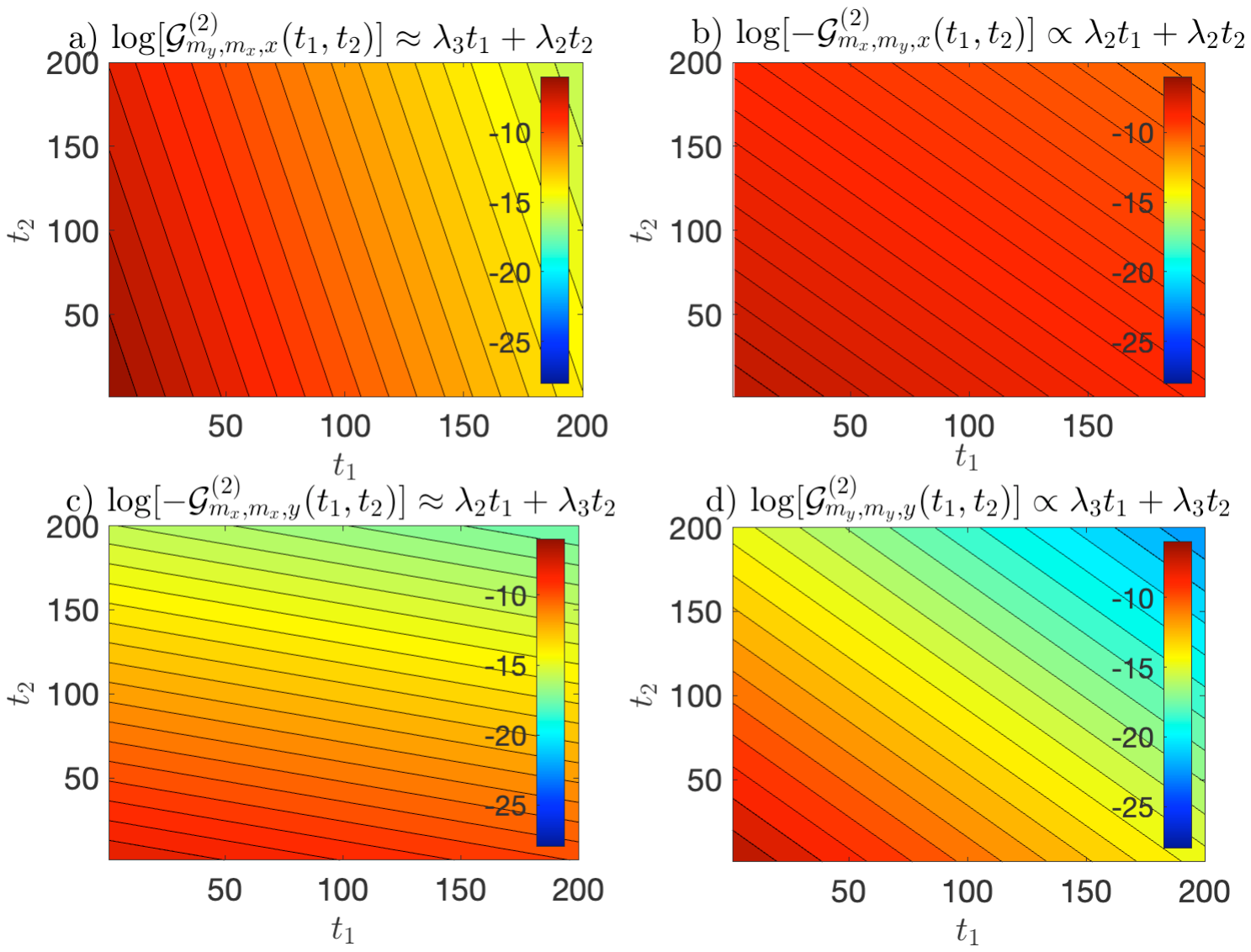}\\
    \caption{Second order Green's functions. Note that decay rates are controlled by a suitable combination of the two eigenvalues of the Koopman operator. Results obtained using RMM. See Eqs. \ref{secondorderresponse}-\ref{Green2mxmy}.}
    \label{GreensRMM}
\end{figure}

\begin{figure}
    \centering
\includegraphics[width=0.9\textwidth]{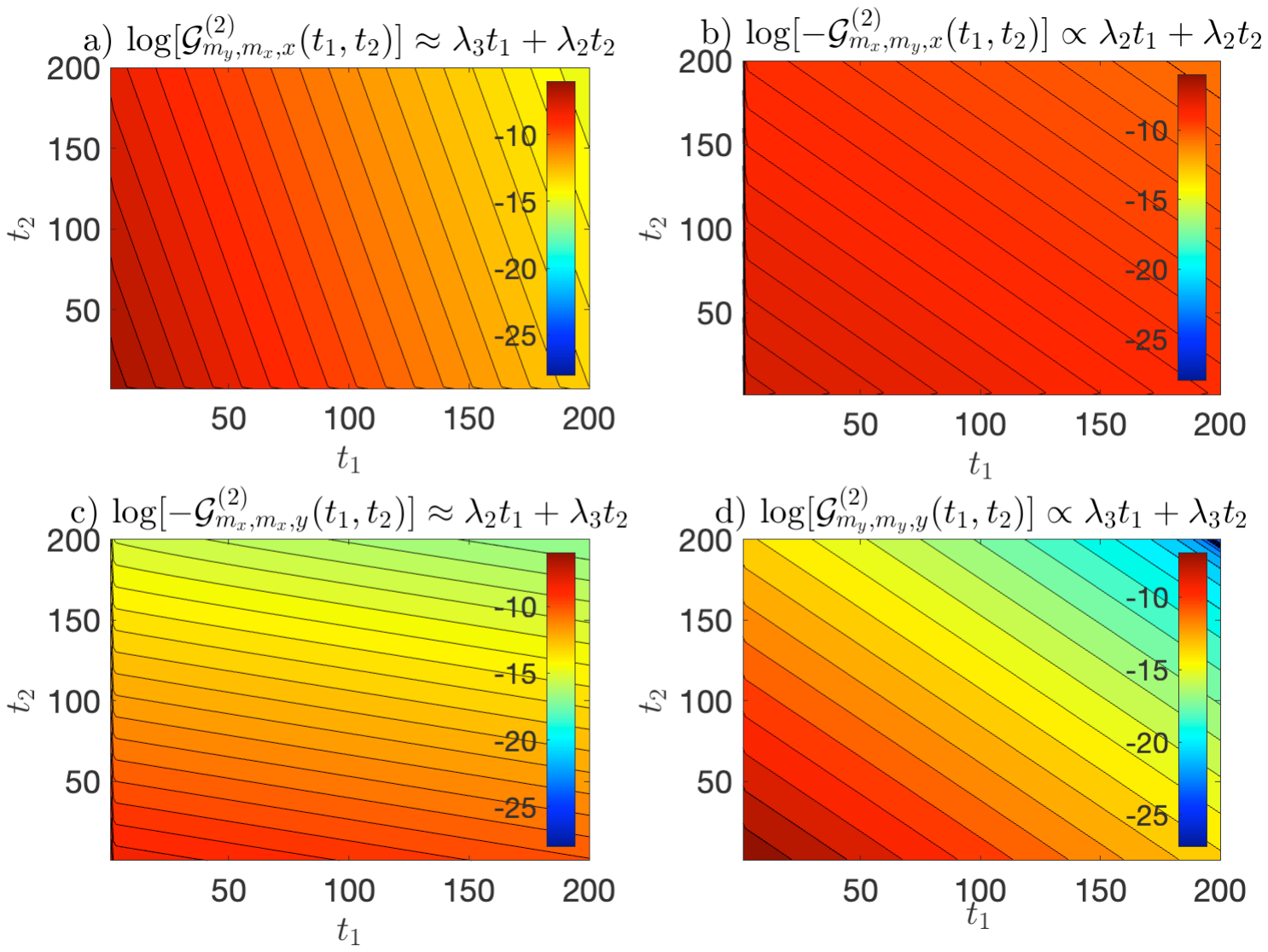}\\
    \caption{Same as in Fig. \ref{GreensRMM}, but here the the results have been obtained using the Ulam discretization.}
    \label{GreensUlam}
\end{figure}

The non-vanishing second-order Green's functions computed for RMM are reported in Fig. \ref{GreensRMM}. As we see, they follow closely the functional dependence derived in Eq. \ref{decompositionG2}, where in this case we obtain simple monomials because only one of the factors $\alpha_{ij}=\langle m_p^T\Pi_j m_q \Pi_i \Psi,\nu_{inv}\rangle$ is non-vanising, because of symmetry, for a choice $p,q=x,y$; $i,j=2,3$; and $\Psi=x,y$. It is extremely encouraging to observe that, just as for the linear case, if we repeat the same analysis using the high-resolution Ulam discretization the results are basically unchanged, compare Figs. \ref{GreensRMM}-\ref{GreensUlam} despite the presence of hundreds of Koopman modes, and hence of hundreds of thousands of $\alpha_{ij}$ factors, only one monomial appears to contribute to the second order Green's function, thus supporting the efficiency of the model reduction attained with the RMM.

\section{Conclusions}\label{conclusions}\label{conclusions}
Constructing accurate and efficient response operators for complex systems is a problem of both theoretical and practical relevance across multiple fields in quantitative sciences \cite{marconi2008fluctuation,Baiesi2013,ruelle2009,Lucarini2020PRSA}. In the case of systems possessing smooth invariant measures, it is possible to resort to non-standard formulations of the FDT to recover such response operators. By combining such a formalism with Koopmanism \cite{Budisic2012}, one gains the important property of interpretability, as it is possible to decompose the Green's functions of interest into a sum of terms, each associated with a specific mode of variability of the system \cite{Santos2022,Zagli2024}. The use of Koopmanism is instrumental for establishing response formulas valid also in the case the stochastic component of a system includes jump processes \cite{Chekroun2024Kolmogorov}.

A key problem in the use of response theory is that one usually needs full knowledge of the evolution equations in order to construct the Green functions and translate them in usable objects at algorithmic level. The latter task is extremely daunting especially when one deals with systems obeying deterministic evolution laws \cite{Wang2013,Chandramoorthy2020,Ni2023,Ni2024}. Recently, it has been proposed to derive response operators by deploying fairly sophisticated machine learning methods based on generative score model \cite{Giorgini2024}. Such a strategy, despite its great potential, is not a silver bullet for cracking the problem of constructing response operators for high dimensional models because of the need to train properly and extensively the surrogate model and issues with out-of-sample performance.  

It is indeed possible to devise workarounds to derive Green's functions even in the case of extremely high dimensional systems by performing suitably defined perturbation experiments, as done in the case of climate models, where the direct evaluation of response operators seems an insurmountable task \cite{Ragone2016,Lucarini2017,Lembo2020}. The flip side is that one can indeed construct useful and accurate black-box-like objects that translate forcings into predicted response, but lack the ability to disentangle and possibly organise hierarchically the impact of the multiple ongoing physical processes. Hence, the level of interpretability of the response operators, despite their skills, can again be disappointing.

\subsection{Towards  Equation-Free and Interpretable Response Formulas} Here we have developed a rather comprehensive set of response formulas - linear and nonlinear, for observables and for correlations, for static as well as for general time dependent perturbations - for Markov chains possessing a unique ergodic invariant measure, which are of  primary importance for dynamical systems theory and statistical mechanics as a whole. See Appendix \ref{arbitrary} and Appendix \ref{responsedynamiccorrelation} for the  general formulas. Appendix \ref{ep} presents a linear response formulas for the total entropy production of a Markov chain undergoing time-dependent perturbation. All of the matrix equations reported in this paper are pseudo-codes that can be seamlessly translated into functioning routines using e.g. software like MATLAB \cite{MATLAB2024} (used here by the author) or open source equivalents like Julia \cite{JULIA}, Python \cite{PYTHON}, and Octave \cite{OCTAVE}, among others.

Hopefully these results can be useful for advancing our understanding of the sensitivity of Markov chains to perturbations and, in particular, of the response near criticality, associated with the closure of the spectral gap of the unperturbed transfer operator \cite{Lucarini2016,Tantet2018,Tantet2018b,Chek_al14_RP,Chekroun_al_RP2,Santos2022}. We will focus on this specific and extremely important problem in a separate study. As for future investigations, it is also tempting to explore the case of absorbing Markov chains, which describe processes where there is a hole (or a trap) in the reference state space, so that the process is eventually killed  \cite{collet2013quasi}. In many cases it is possible to define and prove the existence of quasi-stationary and quasi-ergodic measures, which are constructed by adapting the usual notions to this specific case where one needs to take into account of the continuous leak of mass occurring in the state space \cite{Castro2024}. Since the existence of such measures requires, roughly speaking, the presence of a spectral gap of the transfer operator, it seems interesting to explore whether response formulas could be developed also for such Markov chains.

Going back to applications, the main idea of this paper is to delineate a methodological pipeline for developing simple response formulas that 
\begin{enumerate}
    \item can be  used in a purely data-driven environment, or even if we do not know the evolution equations of the system;
    \item can be cast as simple algebraic operations performed with matrices, thus taking full advantage of the outstanding development occurred in the last decades in numerical linear algebra and the vast availability of dedicated software environments;
    \item allow for a clear interpretation of the response operator thanks to the use of Koopman formalism in the finite state space of  the reduced order model.  
\end{enumerate} 
We have thoroughly explored the efficacy of our strategy on a simple 2D gradient flow forced by additive and diagonal gaussian noise. The system we have investigated features three competing metastable states. We have been able to construct linear and nonlinear response operators that have allow us to define sensitivity and explore response to time-dependent perturbations for observables as well as for correlations of the system. We have shown that even considering a very severely reduced discrete representation of the system, we are able to obtain high-quality information on its response to perturbations and to associated the response to specific modes of unperturbed variability.

The framework above foresees the use of suitable methods of reduction of complexity of a system via MSM before the response theory developed here is used. MSM is very effective in creating a surrogate representation of a possibly multiscale, many-particle system in a moderate number of states \cite{Husic2018}. {\color{black}What we propose here is to extend MSM in such a way that fairly general perturbations can be dealt with.} All one needs is a reference dataset plus few extra datasets produced with a slightly perturbed dynamics, thus allowing to obtain an estimate of the unperturbed Perron-Frobenius operator and of its perturbation in the basis defined by the MSM. Indeed, one can then study the response of the system directly at the desired coarse grained level, bypassing the need to look into all the intricacies of the  system in the original resolution.
The wide availability of software tools facilitating the construction of MSM  \cite{Senne2012,Scherer2015,Harrigan2017}, the wide range of areas of applications for MSM \cite{Husic2018,Pande2010}, recently extended to also to climate applications \cite{Springeretal2024}, and the growing evidence of the efficacy of MSM in capturing the correlation properties of the full system \cite{Suarez2021} support our strategy. 

As mentioned above, constructing a discrete-time, finite-state Markov chain representation of a system amounts to applying the Koopman formalism where the elements of the dictionary are the characteristic functions of the Vorono\"i cells of the phase space. Hence, the results presented here can be seen as linking our attempts at combining of response theory and Koopmanism \cite{Santos2022,Zagli2024,Lucarinietal2025} with the multiplicative DMD algorithm \cite{Boulle2024}. The multiplicative DMD could be improved by using bisecting k-means to create the centroids of the Vorono\"i tessellation, as this methodology is better suited for detecting multiscale features and adapts well to high-dimensional datasets \cite{Souza2024}. {\color{black} Increased accuracy in the estimate of response operators via Koopman methods is likely to be obtained by adopting more sophisticated DMD methods that avoid the problem of spectral pollution \cite{ColbrookTownsend2024}.

Concluding, it is worth noting that in many real-life applications, markovianity is a convenient, useful lie and the lack thereof is the honest, uncomfortable truth. Hence, in the future efforts should be directed at extending the results presented here to the case of Markov chains with memory \cite{Wu2017} space and hidden Markov models \cite{Mor2021}.}


\subsection{A Final Comment on the Prony Method}
In \cite{LucariniChekroun2023,LucariniChekroun2024} we explored the fact that response theory provides solid foundations for a key statistical analysis method used in climate  science, the optimal fingerprinting method for detection and attribution of climate change \cite{Hasselmann1997,Allen1999,Hegerl2011,Hannart2014}. Also the results presented in this contribution seem to provide some clarification for a widely used statistical analysis method used, in this case, in many signal processing applications. In Eqs. \ref{Gdecompo}-\ref{KoopmanResponseb} we have shown that the Koopman operator-based expansion of the linear Green's functions allows to decomposite it in a sum of exponential terms with weighting factors that depend on the chosen observables and on the applied forcings, whilst the exponential decay rates depend exclusively on the properties of the unperturbed system. Indeed, this functional representation points directly to the popular Prony method, which aims at representing the (in general, multivariate) response collected at discrete times of a general system to instantaneous perturbations as a weighted sum of exponentials \cite{Hua1990,Park1999,Kunis2016,Rodriguez2018}. Usually, the number of exponentials one needs to use is a free and uncertain metaparameter of the statistical method. Our approach provides an interpretation of such a metaparameter, which corresponds to the number of discrete states (plus one) we consider in a hypothetical Markov state representation of the system. The well-known uncertainty in defining the optimal value for the metaparameter in the presence of strong noise and/or limited amount of available data can be linked to the difficulty in constructing an accurate Markov model in such conditions. {\color{black}Our observation is \textit{de facto} dual to the findings presented in \cite{Susuki2015}, where it was shown that a multivariate version of the Prony method can be used to perform DMD analysis of a nonlinear dynamical system. }

{\color{black}\ack{VL wishes to thank M. Colbrook, I. Mezi\'c, A. Souza, and N. Zagli and three anonymous reviewers for having provided very insightful comments on the first version  of this paper.}}

\dataccess{
The data needed to reproduce the figures included in the paper and the key components of the code are available here: \texttt{https://figshare.com/s/60cc002c081fdcbadabc}.}

\competing{The author declares no competing interests.}

\funding{This research has been partially supported by the Horizon Europe Project ClimTIP (Grant No. 100018693), {\color{black}by the ARIA SCOP-PR01-P003 - Advancing Tipping Point Early Warning AdvanTip project, by the European Space Agency project PREDICT (contract 4000146344/24/I-LR),} by the EPSRC project LINK (Grant No. EP/Y026675/1), and by the CROPS RETF project funded by the University of Reading.}

\appendix
\section{General formulas for arbitrary order Green's function}\label{arbitrary}
It is relatively straightforward to prove by induction the following result for the $j^{th}$ order response of a generic observable $\Psi$. We have:
\begin{align}
    \frac{1}{n!}&\frac{\mathrm{d^n}\langle \Psi,\nu(n) \rangle}{\mathrm{d}\epsilon^n}\big{|}_{\epsilon=0}=\langle \Psi,\nu^{(j)}(n) \rangle\nonumber\\
    &= \sum_{k_1=-\infty}^\infty \Theta(k_1) \langle m^T  (\mathcal{M}^T)^k  \Psi,\nu^{(j-1)}(n-k_1-1)\rangle f(n-k_1-1)\nonumber\\
    &=\sum_{k_1,\ldots,k_j=-\infty}^\infty\Theta(k_1)\ldots\Theta(k_n) \langle m^T(\mathcal{M}^T)^{k_n}\ldots m^T(\mathcal{M}^T)^{k_1} \Psi ,\nu_{inv}\rangle \times\nonumber \\&f(n-\sum_{p=1}^jk_p-j)\ldots f(n-k_1-1)\nonumber\\
    & =  (\mathcal{G}^{(n)}_{m,\Psi}\star f)(n)  
\end{align}
where $\mathcal{G}^{(n)}_{m,\Psi}(k_1,\ldots,k_j)=\Theta(k_1)\ldots\Theta(k_n)\langle m^T(\mathcal{M}^T)^{k_n}\ldots m^T(\mathcal{M}^T)^{k_1} \Psi ,\nu_{inv}\rangle$ and where $*$ indicates here a n-uple convolution sum. 

Using the spectral expansion of the $\mathcal{M}$ operator in Kolmogorov modes, we have:

\begin{align}
    \mathcal{G}^{(n)}_{m,\Psi}(k_1,\ldots,k_n)=\Theta(k_1)\ldots\Theta(k_n)\sum_{i_1,\ldots i_n=2}^N\alpha_{i_1,\ldots,i_n}\lambda_{i_1}^{k_1}\ldots \lambda_{i_n}^{k_n}, \quad \alpha_{i_1,\ldots,i_n}=\langle m^T \Pi_{i_n} \ldots m^T\Pi_{i_1} \Psi ,\nu_{inv}\rangle.
\end{align}
Hence, we derive that the Green's function of order $n$ is a function of $n$ variables that can be written as a sum of exponentials that decrease with time, considering that $\lambda_i^k=\exp(k\nu_i)$, with $\nu_i=\log(\lambda_i)<0$.  Note that the previous result provides also a general statistical model one can use to fit experimental or model generated data.

\section{Dynamic Response of Correlations}\label{responsedynamiccorrelation}
We extend here the results presented in Sect. \ref{responsecorrelations} to the more general case $\mathcal M\rightarrow \mathcal{M}_{\epsilon,n}=\mathcal{M}+\epsilon f(n)m$. Since the system has explicit time-dependent dynamics, so that the statistical properties are indexed by the observation time $n$. We have:
\begin{align}
    C^\epsilon_{n,l}(\Psi,\Phi)&=   \langle \prod_{p=0}^{l-1} (\mathcal{M}^T+\epsilon f(n+p)m^T)^l\Psi \circ \Phi,\nu_{inv}+\sum_{p=1}^\infty\epsilon^p\nu^{(p)}(n)\rangle \nonumber\\&-\langle \Psi,\nu_\epsilon(n)\rangle \langle \Phi,\nu_\epsilon(n)\rangle   
\end{align}
If we now collect the terms up to first order in $\epsilon$, we obtain:
\begin{align}
    &  \langle \prod_{q=0}^{l-1} (\mathcal{M}^T+\epsilon m^T f(n+q))\Psi \circ \Phi,\nu_{inv}+\sum_{p=1}^\infty\epsilon^p\nu^{(p)}(n)\rangle   \nonumber \\
+    &\langle ( \mathcal{M}^T)^l\Psi \circ \Phi,\nu_{inv}\rangle \nonumber  \\
     & + \epsilon  \sum_{q=0}^{l-1} \langle (\mathcal{M}^{l-q-1})^T m^T f(n+q) (\mathcal{M}^q)^ \Psi\circ\Phi,\nu_{inv}\rangle \nonumber\\
     & + \epsilon \sum_{k=-\infty}^\infty \langle (\mathcal{M}^T)^l\Psi \circ \Phi, \Theta(k)  \mathcal{M}^k  m \nu_{inv}f(n-k-1)\rangle  +  o(\epsilon)
\end{align}
Additionally, up to first order in $\epsilon$, we have
\begin{align}
&\langle \Psi,\nu_{inv}+\sum_{p=1}^\infty\epsilon^p\nu^{(p)}(n)\rangle \langle\Phi,\nu_{inv}+\sum_{p=1}^\infty\epsilon^p\nu^{(p)}(n)\rangle \rangle\nonumber\\=
&\langle \Psi,\nu_{inv}\rangle \langle\Phi,\nu_{inv}\rangle\nonumber\\+
&\epsilon \sum_{k=-\infty}^\infty \langle \Psi,\Theta(k) \mathcal{M}^k  m \nu_{inv}f(n-k-1) \rangle\langle\Phi,\nu_{inv}\rangle\nonumber\\
&+\epsilon \langle\Psi,\nu_{inv}\rangle\sum_{k=-\infty}^\infty \langle \Phi, \Theta(k)\mathcal{M}^k  m \nu_{inv}f(n-k-1) \rangle +o(\epsilon)
\end{align}
We derive our final result:
\begin{align}
\frac{\mathrm{d}C^\epsilon_{l,n}(\Psi,\Phi)}{\mathrm{d}\epsilon}|_{\epsilon=0}&= \sum_{q=0}^{l-1} \langle (\mathcal{M}^{l-q-1})^T m^T f(n+q) (\mathcal{M}^{q})^T\Psi\circ\Phi,\nu_{inv}\rangle\nonumber \\
& \sum_{k=-\infty}^\infty \Theta(k) \langle m^T(\mathcal{M}^k)^T((\mathcal{M}^{l})^T\Psi\circ\Phi-\Psi \langle\Phi,\nu_{inv}\rangle - \Phi\langle\Psi),\nu_{inv}\rangle),  \nu_{inv}\rangle f(n-k-1), 
\end{align}
which clearly agrees with the case of static perturbation shown before if one assumes $f=1$ and $n\rightarrow\infty$. 

\section{Entropy Production}\label{ep}
The total entropy production at time $n$ for a time-dependent Markov chain can be written as:
\begin{align}
    S_{tot}(n)&= \sum_{i,j} \mathcal{M}_\epsilon(n)_{ij}\nu(n)_j \ln \left( \frac{\mathcal{M}_\epsilon(n)_{ij}\nu(n)_j}{\mathcal{M}_\epsilon(n)_{ji}\nu(n)_i} \right)\nonumber\\
    &=\frac{1}{2}\sum_{i,j} (\mathcal{M}_\epsilon(n)_{ij}\nu(n)_j-\mathcal{M}_\epsilon(n)_{ji}\nu(n)_i) \ln \left( \frac{\mathcal{M}_\epsilon(n)_{ij}\nu(n)_j}{\mathcal{M}_\epsilon(n)_{ji}\nu(n)_i} \right)
\end{align}
where we have adapted Seifert's results \cite{Seifert2005} to the case of discrete time dynamics. We want to expand the previous expression up to first order in $\epsilon$. We obtain after lengthy but straightforward calculations:
\begin{equation}
    S_{tot}(n)=S^{(0)}_{tot}+\frac{\partial  S_{tot}(n)}{\partial \epsilon}|_{\epsilon=0}\epsilon+h.o.t.\end{equation}
    where
 \begin{equation}
  S^{(0)}_{tot}= \sum_{i,j} \mathcal{M}_{ij}(\nu_{inv})_j \ln \left( \frac{\mathcal{M}_{ij}(\nu_{inv})_j}{\mathcal{M}_{ji}(\nu_{inv})_i} \right)
\end{equation}
and the linear response formula for the entropy production is given by
\begin{equation}
    \frac{\partial  S_{tot}(n)}{\partial \epsilon}|_{\epsilon=0}= \sum_{i,j} \left(m_{ij}f(n)(\nu_{inv})_j+\mathcal{M}_{ij} \nu^{(1)}_j(n)\right)\ln \left( \frac{\mathcal{M}_{ij}(\nu_{inv})_j}{\mathcal{M}_{ji}(\nu_{inv})_i}, \right)
\end{equation}
where, following Eq. \ref{deltanu1}
\begin{equation}
   \nu^{(1)}_j(n)=\sum_{k=-\infty}^\infty \sum_i\Theta(k)  (\mathcal{M}^k  m)_{ji} (\nu_{inv})_if(n-k-1).
\end{equation}
Note that if the unperturbed system obeys detailed balance ($\mathcal{M}_{ji}(\nu_{inv})_i=\mathcal{M}_{ij}(\nu_{inv})_j$ $\forall i,j$), the entropy production of the unperturbed state as well as its sensitivity with respect to $\epsilon$ vanish.




\providecommand{\noopsort}[1]{}\providecommand{\singleletter}[1]{#1}%

\vskip2pc


\begin{thebibliography}{99}

\bibitem{Hanggi1982}
H\"anggi P, Thomas H. 1982  Stochastic processes: Time evolution, symmetries
  and linear response. {\em Physics Reports} \textbf{88}, 207 -- 319.
(\href{http://dx.doi.org/https://doi.org/10.1016/0370-1573(82)90045-X}{https://doi.org/10.1016/0370-1573(82)90045-X})

\bibitem{Baiesi2013}
Baiesi M, Maes C. 2013  An update on the nonequilibrium linear response. {\em
  New Journal of Physics} \textbf{15}, 013004.

\bibitem{Sarracino2019}
Sarracino A, Vulpiani A. 2019  On the fluctuation-dissipation relation in
  non-equilibrium and non-Hamiltonian systems. {\em Chaos} \textbf{29}, 083132.

\bibitem{Kubo1966}
Kubo R. 1966  {The fluctuation-dissipation theorem}. {\em Reports on Progress
  in Physics} \textbf{29}, 255--284.

\bibitem{Hairer2010}
Hairer M, Majda AJ. 2010  A simple framework to justify linear response theory.
  {\em Nonlinearity} \textbf{23}, 909--922.
(\href{http://dx.doi.org/10.1088/0951-7715/23/4/008}{10.1088/0951-7715/23/4/008})

\bibitem{marconi2008fluctuation}
Marconi UMB, Puglisi A, Rondoni L, Vulpiani A. 2008  {Fluctuation--dissipation:
  response theory in statistical physics}. {\em Physics reports} \textbf{461},
  111--195.

\bibitem{pavliotisbook2014}
Pavliotis GA. 2014 {\em {Stochastic Processes and Applications}} vol.~60.
Springer, New York.
(\href{http://dx.doi.org/10.1007/978-1-4939-1323-7}{10.1007/978-1-4939-1323-7})

\bibitem{ruellegeneral1998}
Ruelle D. 1998  {General linear response formula in statistical mechanics, and
  the fluctuation-dissipation theorem far from equilibrium}. {\em Physics
  Letters, Section A: General, Atomic and Solid State Physics} \textbf{245},
  220--224.
(\href{http://dx.doi.org/10.1016/S0375-9601(98)00419-8}{10.1016/S0375-9601(98)00419-8})

\bibitem{ruelle2009}
Ruelle D. 2009  {A review of linear response theory for general differentiable
  dynamical systems}. {\em Nonlinearity} \textbf{22}, 855--870.

\bibitem{Liverani2006}
Liverani C, Gou\"ezel S. 2006  {Banach spaces adapted to Anosov systems}. {\em
  Ergodic Theory and Dynamical Systems} \textbf{26}, 189--217.

\bibitem{Baladi2014}
Baladi V. 2014  {Linear response, or else}. {\em Proceedings of the
  International Congress of Mathematicians-Seoul} \textbf{3}, 525--545.

\bibitem{Baladi2017a}
Baladi V, Kuna T, Lucarini V. 2017  {Linear and fractional response for the SRB
  measure of smooth hyperbolic attractors and discontinuous observables}. {\em
  Nonlinearity} \textbf{30}, 1204--1220.
(\href{http://dx.doi.org/10.1088/1361-6544/aa5b13}{10.1088/1361-6544/aa5b13})

\bibitem{ruelle_nonequilibrium_1998}
Ruelle D. 1998  Nonequilibrium statistical mechanics near equilibrium:
  computing higher-order terms. {\em Nonlinearity} \textbf{11}, 5--18.

\bibitem{lucarini2008}
Lucarini V. 2008  {Response theory for equilibrium and non-equilibrium
  statistical mechanics: Causality and generalized kramers-kronig relations}.
  {\em Journal of Statistical Physics} \textbf{131}, 543--558.
(\href{http://dx.doi.org/10.1007/s10955-008-9498-y}{10.1007/s10955-008-9498-y})

\bibitem{Lucarini2009DispersionNonLinearResponseLorenz}
Lucarini V. 2009  Evidence of dispersion relations for the nonlinear response
  of the {L}orenz 63 system. {\em J. Stat. Phys.} \textbf{134}, 381--400.
(\href{http://dx.doi.org/10.1007/s10955-008-9675-z}{10.1007/s10955-008-9675-z})

\bibitem{Chekroun_al_RP2}
Chekroun M, Tantet A, Dijkstra H, Neelin JD. 2020  {Ruelle-Pollicott Resonances
  of Stochastic Systems in Reduced State Space. Part I: Theory}. {\em
  J.~Stat.~Phys.} \textbf{179}, 1366--1402.

\bibitem{Santos2022}
Santos~Guti{\'e}rrez M, Lucarini V. 2022  On some aspects of the response to
  stochastic and deterministic forcings. {\em Journal of Physics A:
  Mathematical and Theoretical} \textbf{55}, 425002.
(\href{http://dx.doi.org/10.1088/1751-8121/ac90fd}{10.1088/1751-8121/ac90fd})

\bibitem{LC2023}
Lucarini V, Chekroun MD. 2023  Theoretical tools for understanding the climate
  crisis from Hasselmann's programme and beyond. {\em Nature Reviews Physics}.
(\href{http://dx.doi.org/10.1038/s42254-023-00650-8}{10.1038/s42254-023-00650-8})

\bibitem{Mezic2005}
Mezić I. 2005  Spectral Properties of Dynamical Systems, Model Reduction and
  Decompositions. {\em Nonlinear Dynamics} \textbf{41}, 309--325.
(\href{http://dx.doi.org/10.1007/s11071-005-2824-x}{10.1007/s11071-005-2824-x})

\bibitem{Budisic2012}
Budi\v{s}i\'{c} M, Mohr R, Mezi\'{c} I. 2012  Applied {K}oopmanism. {\em Chaos}
  \textbf{22}, 047510, 33.
(\href{http://dx.doi.org/10.1063/1.4772195}{10.1063/1.4772195})

\bibitem{Kutz2016}
Kutz J, Brunton S, Brunton B, Proctor J. 2016 {\em Dynamic Mode Decomposition:
  Data-Driven Modeling of Complex Systems}.
Other Titles in Applied Mathematics. Philadelphia: Society for Industrial and
  Applied Mathematics.

\bibitem{brunton2022data}
Brunton SL, Kutz JN. 2022 {\em {Data-driven Science and Engineering: Machine
  Learning, Dynamical Systems, and Control}}.
Cambridge University Press.

\bibitem{Klus2018}
Klus S, N\"{u}ske F, Koltai P, Wu H, Kevrekidis I, Sch\"{u}tte C, No\'{e} F.
  2018  Data-driven model reduction and transfer operator approximation. {\em
  J. Nonlinear Sci.} \textbf{28}, 985--1010.
(\href{http://dx.doi.org/10.1007/s00332-017-9437-7}{10.1007/s00332-017-9437-7})

\bibitem{Colbrook2024Multi}
Colbrook MJ. 2024  Chapter 4 - The multiverse of dynamic mode decomposition
  algorithms. In Mishra S, Townsend A, editors, {\em Numerical Analysis Meets
  Machine Learning}, Handbook of Numerical Analysis,  vol.~25,  pp. 127--230.
  Elsevier.
(\href{http://dx.doi.org/https://doi.org/10.1016/bs.hna.2024.05.004}{https://doi.org/10.1016/bs.hna.2024.05.004})

\bibitem{Zagli2024}
Zagli N, Colbrook M, Lucarini V, Mezić I, Moroney J. 2024  Bridging the Gap
  between Koopmanism and Response Theory: Using Natural Variability to Predict
  Forced Response. {\em arXiv:2410.01622}.
(\href{http://dx.doi.org/https://arxiv.org/abs/2410.01622}{https://arxiv.org/abs/2410.01622})

\bibitem{Lucarinietal2025}
Lucarini V, Gutierrez MS, Moroney J, Zagli N. 2025  General Framework for
  Linking Free and Forced Fluctuations via Koopmanism. {\em arXiv:2506.16446}.
(\href{http://dx.doi.org/https://arxiv.org/abs/arXiv:2506.16446}{https://arxiv.org/abs/arXiv:2506.16446})

\bibitem{Chekroun2024Kolmogorov}
Chekroun MD, Zagli N, Lucarini V. 2024  Kolmogorov Modes and Linear Response of
  Jump-Diffusion Models: Applications to Stochastic Excitation of the ENSO
  Recharge Oscillator. {\em arXiv preprint arXiv:2411.14769}.

\bibitem{Norris1998}
Norris JR. 1998 {\em Markov Chains}.
Cambridge University Press.

\bibitem{behrends2000introduction}
Behrends E. 2000 {\em Introduction to Markov Chains}.
Vieweg+Teubner Verlag.
(\href{http://dx.doi.org/10.1007/978-3-663-11603-3}{10.1007/978-3-663-11603-3})

\bibitem{bowen1970markov}
Bowen R. 1970  Markov Partitions for Axiom A Diffeomorphisms. {\em American
  Journal of Mathematics} \textbf{92}, 725--747.
(\href{http://dx.doi.org/10.2307/2373370}{10.2307/2373370})

\bibitem{attal2010markov}
Attal S. 2010  Markov Chains and Dynamical Systems: The Open System Point of
  View. {\em Communications on Stochastic Analysis} \textbf{4}, 447--466.
(\href{http://dx.doi.org/10.31390/cosa.4.4.05}{10.31390/cosa.4.4.05})

\bibitem{Ulam1960}
Ulam S. 1960 {\em A Collection of Mathematical Problems}.
Interscience tracts in pure and applied mathematics. Interscience Publishers.

\bibitem{baladi2000positive}
Baladi V. 2000 {\em {Positive Transfer Operators and Decay of Correlations}}
  vol.~16.
World scientific.

\bibitem{Froyland1998}
Froyland G. 1998  Approximating physical invariant measures of mixing dynamical
  systems in higher dimensions. {\em Nonlinear Analysis: Theory, Methods \&
  Applications} \textbf{32}, 831--860.
(\href{http://dx.doi.org/https://doi.org/10.1016/S0362-546X(97)00527-0}{https://doi.org/10.1016/S0362-546X(97)00527-0})

\bibitem{Ding2002}
Ding J, Li TY, Zhou A. 2002  Finite approximations of Markov operators. {\em
  Journal of Computational and Applied Mathematics} \textbf{147}, 137--152.
(\href{http://dx.doi.org/10.1016/S0377-0427(02)00429-6}{10.1016/S0377-0427(02)00429-6})

\bibitem{Aurenhammer1991}
Aurenhammer F. 1991  Voronoi diagrams—a survey of a fundamental geometric
  data structure. {\em ACM Comput. Surv.} \textbf{23}, 345–405.
(\href{http://dx.doi.org/10.1145/116873.116880}{10.1145/116873.116880})

\bibitem{Forgy65}
Forgy E. 1965  Cluster Analysis of Multivariate Data: Efficiency versus
  Interpretability of Classification. {\em Biometrics} \textbf{21}, 768--769.

\bibitem{Lloyd82}
Lloyd S. 1982  Least squares quantization in PCM. {\em IEEE Transactions on
  Information Theory} \textbf{28}, 129--137.
(\href{http://dx.doi.org/10.1109/TIT.1982.1056489}{10.1109/TIT.1982.1056489})

\bibitem{Laio2006}
Bussi G, Laio A, Parrinello M. 2006  Equilibrium Free Energies from
  Nonequilibrium Metadynamics. {\em Phys. Rev. Lett.} \textbf{96}, 090601.
(\href{http://dx.doi.org/10.1103/PhysRevLett.96.090601}{10.1103/PhysRevLett.96.090601})

\bibitem{Pande2010}
Pande VS, Beauchamp K, Bowman GR. 2010  Everything you wanted to know about
  Markov State Models but were afraid to ask. {\em Methods} \textbf{52},
  99--105.
(\href{http://dx.doi.org/10.1016/j.ymeth.2010.06.002}{10.1016/j.ymeth.2010.06.002})

\bibitem{Bowman2014}
Bowman GR, Pande V, Noé F. 2014 {\em An introduction to Markov State Models
  and their application to long timescale molecular simulation}.
Advances in Experimental Medicine and Biology, volume 797. Dordrecht: Springer.

\bibitem{Husic2018}
Husic BE, Pande VS. 2018  Markov State Models: From an Art to a Science. {\em
  Journal of the American Chemical Society} \textbf{140}, 2386--2396.
doi: 10.1021/jacs.7b12191
  (\href{http://dx.doi.org/10.1021/jacs.7b12191}{10.1021/jacs.7b12191})

\bibitem{Froyland2014}
Froyland G, Gottwald GA, Hammerlindl A. 2014  A Computational Method to Extract
  Macroscopic Variables and Their Dynamics in Multiscale Systems. {\em SIAM
  Journal on Applied Dynamical Systems} \textbf{13}, 1816--1846.
(\href{http://dx.doi.org/10.1137/130943637}{10.1137/130943637})

\bibitem{Bittracher2018}
Bittracher A, Koltai P, Klus S, Banisch R, Dellnitz M, Schütte C. 2018
  Transition Manifolds of Complex Metastable Systems. {\em Journal of Nonlinear
  Science} \textbf{28}, 471--512.
(\href{http://dx.doi.org/10.1007/s00332-017-9415-0}{10.1007/s00332-017-9415-0})

\bibitem{mori_transport_1965}
Mori H. 1965  Transport, Collective Motion, and {Brownian} Motion. {\em
  Progress of Theoretical Physics} \textbf{33}, 423--455.

\bibitem{zwanzig_memory_1961}
Zwanzig R. 1961  Memory Effects in Irreversible Thermodynamics. {\em Physical
  Review} \textbf{124}, 983--992.

\bibitem{Kalliadasis2015}
Kalliadasis S, Krumscheid S, Pavliotis G. 2015  A new framework for extracting
  coarse-grained models from time series with multiscale structure. {\em
  Journal of Computational Physics} \textbf{296}, 314--328.
(\href{http://dx.doi.org/https://doi.org/10.1016/j.jcp.2015.05.002}{https://doi.org/10.1016/j.jcp.2015.05.002})

\bibitem{Chekroun2015b}
Chekroun MD, Liu H, Wang S. 2015 {\em Stochastic Parameterizing Manifolds and
  Non-Markovian Reduced Equations}.
SpringerBriefs in Mathematics. Cham: Springer International Publishing.

\bibitem{santos2021reduced}
Santos~Guti{\'e}rrez M, Lucarini V, Chekroun MD, Ghil M. 2021  {Reduced-order
  models for coupled dynamical systems: Data-driven methods and the Koopman
  operator}. {\em Chaos} \textbf{31}, 053116.

\bibitem{Chekroun2021}
Chekroun MD, Liu H, McWilliams JC. 2021  Stochastic rectification of fast
  oscillations on slow manifold closures. {\em Proc. Natl. Acad. Sci. USA}
  \textbf{118}, Paper No. e2113650118, 9.
(\href{http://dx.doi.org/10.1073/pnas.2113650118}{10.1073/pnas.2113650118})

\bibitem{Chekroun2025}
Chekroun MD, Liu H, McWilliams JC. 2025  Non-Markovian reduced models to
  unravel transitions in non-equilibrium systems. {\em Journal of Physics A:
  Mathematical and Theoretical} \textbf{58}, 045204.
(\href{http://dx.doi.org/10.1088/1751-8121/ada7ad}{10.1088/1751-8121/ada7ad})

\bibitem{Gear2003}
Kevrekidis IG, Gear CW, Hyman JM, Kevrekidis PG, , Runborg O, Theodoropoulos C.
  2003  Equation-Free, Coarse-Grained Multiscale Computation: Enabling
  Macroscopic Simulators to Perform System-Level Analysis. {\em Communications
  in Mathematical Sciences} \textbf{1}, 718--762.
(\href{http://dx.doi.org/10.4310/CMS.2003.v1.n4.a5}{10.4310/CMS.2003.v1.n4.a5})

\bibitem{Kevrekidis2009}
Kevrekidis IG, Samaey G. 2009  Equation-Free Multiscale Computation: Algorithms
  and Applications. {\em Annual Review of Physical Chemistry} \textbf{60},
  321--344.
(\href{http://dx.doi.org/https://doi.org/10.1146/annurev.physchem.59.032607.093610}{https://doi.org/10.1146/annurev.physchem.59.032607.093610})

\bibitem{Ma2005}
Ma A, Dinner AR. 2005  Automatic Method for Identifying Reaction Coordinates in
  Complex Systems. {\em The Journal of Physical Chemistry B} \textbf{109},
  6769--6779.
(\href{http://dx.doi.org/10.1021/jp045546c}{10.1021/jp045546c})

\bibitem{Rogal2021}
Rogal J. 2021  Reaction coordinates in complex systems-a perspective. {\em The
  European Physical Journal B} \textbf{94}, 223.
(\href{http://dx.doi.org/10.1140/epjb/s10051-021-00233-5}{10.1140/epjb/s10051-021-00233-5})

\bibitem{ZagliLucariniPavliotis}
Zagli N, Lucarini V, Pavliotis GA. 2021  Spectroscopy of phase transitions for
  multiagent systems. {\em Chaos: An Interdisciplinary Journal of Nonlinear
  Science} \textbf{31}, 061103.
(\href{http://dx.doi.org/10.1063/5.0053558}{10.1063/5.0053558})

\bibitem{ZaglietalJPA2024}
Zagli N, Lucarini V, Pavliotis GA. 2024  Response theory identifies reaction
  coordinates and explains critical phenomena in noisy interacting systems.
  {\em J. Phys. A} \textbf{34}, 325004.
(\href{http://dx.doi.org/10.1088/1751-8121/ad6068}{10.1088/1751-8121/ad6068})

\bibitem{Boulle2024}
Boull\'e' N, Colbrook MJ. 2024  Multiplicative Dynamic Mode Decomposition. {\em
  arXiv:2405.05334}.
(\href{http://dx.doi.org/https://arxiv.org/abs/2405.05334}{https://arxiv.org/abs/2405.05334})

\bibitem{Lucarini2016}
Lucarini V. 2016  {Response Operators for Markov Processes in a Finite State
  Space: Radius of Convergence and Link to the Response Theory for Axiom A
  Systems}. {\em Journal of Statistical Physics} \textbf{162}, 312--333.
(\href{http://dx.doi.org/10.1007/s10955-015-1409-4}{10.1007/s10955-015-1409-4})

\bibitem{SantosJSP}
Santos~Guti{\'{e}}rrez M, Lucarini V. 2020  {Response and Sensitivity Using
  Markov Chains}. {\em Journal of Statistical Physics} \textbf{179},
  1572--1593.
(\href{http://dx.doi.org/10.1007/s10955-020-02504-4}{10.1007/s10955-020-02504-4})

\bibitem{Esposito2024a}
Aslyamov T, Esposito M. 2024a  Nonequilibrium Response for Markov Jump
  Processes: Exact Results and Tight Bounds. {\em Phys. Rev. Lett.}
  \textbf{132}, 037101.
(\href{http://dx.doi.org/10.1103/PhysRevLett.132.037101}{10.1103/PhysRevLett.132.037101})

\bibitem{EspositoPRL2024b}
Aslyamov T, Esposito M. 2024b  General Theory of Static Response for Markov
  Jump Processes. {\em Phys. Rev. Lett.} \textbf{133}, 107103.
(\href{http://dx.doi.org/10.1103/PhysRevLett.133.107103}{10.1103/PhysRevLett.133.107103})

\bibitem{Zhang2012}
Zhang XJ, Qian H, Qian M. 2012  Stochastic theory of nonequilibrium steady
  states and its applications. Part I. {\em Physics Reports} \textbf{510},
  1--86.
Stochastic Theory of Nonequilibrium Steady States and Its Applications: Part I
  (\href{http://dx.doi.org/https://doi.org/10.1016/j.physrep.2011.09.002}{https://doi.org/10.1016/j.physrep.2011.09.002})

\bibitem{Koltai2018}
Koltai P, Sch\"{u}tte C. 2018  A Multiscale Perturbation Expansion Approach for
  Markov State Modeling of Nonstationary Molecular Dynamics. {\em Multiscale
  Modeling \& Simulation} \textbf{16}, 1455--1485.
(\href{http://dx.doi.org/10.1137/17M1146403}{10.1137/17M1146403})

\bibitem{Falasco2025}
Falasco G, Esposito M. 2025  Macroscopic stochastic thermodynamics. {\em Rev.
  Mod. Phys.} \textbf{97}, 015002.
(\href{http://dx.doi.org/10.1103/RevModPhys.97.015002}{10.1103/RevModPhys.97.015002})

\bibitem{Seifert2005}
Seifert U. 2005  Entropy Production along a Stochastic Trajectory and an
  Integral Fluctuation Theorem. {\em Phys. Rev. Lett.} \textbf{95}, 040602.
(\href{http://dx.doi.org/10.1103/PhysRevLett.95.040602}{10.1103/PhysRevLett.95.040602})

\bibitem{LucariniWouters2017}
Lucarini V, Wouters J. 2017  Response formulae for n-point correlations in
  statistical mechanical systems and application to a problem of coarse
  graining. {\em Journal of Physics A: Mathematical and Theoretical}
  \textbf{50}, 355003.
(\href{http://dx.doi.org/10.1088/1751-8121/aa812c}{10.1088/1751-8121/aa812c})

\bibitem{Klus2016}
Klus S, Koltai P, Schütte C. 2016  On the numerical approximation of the
  Perron-Frobenius and Koopman operator. {\em Journal of Computational
  Dynamics} \textbf{3}, 51--79.
(\href{http://dx.doi.org/10.3934/jcd.2016003}{10.3934/jcd.2016003})

\bibitem{Hua1990}
Hua Y, Sarkar T. 1990  Matrix pencil method for estimating parameters of
  exponentially damped/undamped sinusoids in noise. {\em IEEE Transactions on
  Acoustics, Speech, and Signal Processing} \textbf{38}, 814--824.
(\href{http://dx.doi.org/10.1109/29.56027}{10.1109/29.56027})

\bibitem{Park1999}
Park S, Schapery R. 1999  Methods of interconversion between linear
  viscoelastic material functions. Part I—a numerical method based on Prony
  series. {\em International Journal of Solids and Structures} \textbf{36},
  1653--1675.
(\href{http://dx.doi.org/https://doi.org/10.1016/S0020-7683(98)00055-9}{https://doi.org/10.1016/S0020-7683(98)00055-9})

\bibitem{Kunis2016}
Kunis S, Peter T, R\"omer T, {von der Ohe} U. 2016  A multivariate
  generalization of Prony's method. {\em Linear Algebra and its Applications}
  \textbf{490}, 31--47.
(\href{http://dx.doi.org/https://doi.org/10.1016/j.laa.2015.10.023}{https://doi.org/10.1016/j.laa.2015.10.023})

\bibitem{Rodriguez2018}
Fernández~Rodríguez A, de~Santiago~Rodrigo L, López~Guillén E,
  Rodríguez~Ascariz JM, Miguel~Jiménez JM, Boquete L. 2018  Coding Prony’s
  method in MATLAB and applying it to biomedical signal filtering. {\em BMC
  Bioinformatics} \textbf{19}, 451.
(\href{http://dx.doi.org/10.1186/s12859-018-2473-y}{10.1186/s12859-018-2473-y})

\bibitem{senetanonnegative1973}
Seneta E. 1973 {\em {Non-negative matrices}}.
George Allen and Unwin.

\bibitem{Abbas2016}
Abbas K, Berkhout J, Heidergott B. 2016  A Critical Account of Perturbation
  Analysis of Markov Chains. {\em Markov Processes and Related Fields}
  \textbf{22}, 227--266.

\bibitem{Mitrophanov_2024}
Mitrophanov AY. 2024  The Arsenal of Perturbation Bounds for Finite
  Continuous-Time Markov Chains: A Perspective. {\em Mathematics} \textbf{12}.
(\href{http://dx.doi.org/10.3390/math12111608}{10.3390/math12111608})

\bibitem{GhilLucarini2020}
Ghil M, Lucarini V. 2020  The physics of climate variability and climate
  change. {\em Rev. Mod. Phys.} \textbf{92}, 035002.
(\href{http://dx.doi.org/10.1103/RevModPhys.92.035002}{10.1103/RevModPhys.92.035002})

\bibitem{LucariniChekroun2023}
Lucarini V, Chekroun MD. 2023  Theoretical tools for understanding the climate
  crisis from Hasselmann's programme and beyond. {\em Nature Reviews Physics}
  \textbf{5}, 744--765.
(\href{http://dx.doi.org/10.1038/s42254-023-00650-8}{10.1038/s42254-023-00650-8})

\bibitem{Ghil2008}
Ghil M, Chekroun MD, Simonnet E. 2008  Climate dynamics and fluid mechanics:
  Natural variability and related uncertainties. {\em Physica D} \textbf{237},
  2111--2126.

\bibitem{Chekroun2011}
Chekroun MD, Simonnet E, Ghil M. 2011  Stochastic climate dynamics: {Random
  attractors and time-dependent invariant measures}. {\em Physica D: Nonlinear
  Phenomena} \textbf{240}, 1685--1700.
(\href{http://dx.doi.org/10.1016/j.physd.2011.06.005}{10.1016/j.physd.2011.06.005})

\bibitem{bodai2011}
B{\'{o}}dai T, Karolyi G, T{\'{e}}l T. 2011  {A chaotically driven model
  climate: Extreme events and snapshot attractors}. {\em Nonlinear Processes in
  Geophysics} \textbf{18}, 573--580.
(\href{http://dx.doi.org/10.5194/npg-18-573-2011}{10.5194/npg-18-573-2011})

\bibitem{Tel2020}
T{\'e}l T, B{\'o}dai T, Dr{\'o}tos G, Haszpra T, Herein M, Kasz{\'a}s B, Vincze
  M. 2020  The Theory of Parallel Climate Realizations. {\em Journal of
  Statistical Physics} \textbf{179}, 1496--1530.
(\href{http://dx.doi.org/10.1007/s10955-019-02445-7}{10.1007/s10955-019-02445-7})

\bibitem{Bodai2020}
B\'odai T, Dr\'otos G, Herein M, Lunkeit F, Lucarini V. 2020  The Forced
  Response of the El Niño–Southern Oscillation–Indian Monsoon
  Teleconnection in Ensembles of Earth System Models. {\em Journal of Climate}
  \textbf{33}, 2163 -- 2182.
(\href{http://dx.doi.org/10.1175/JCLI-D-19-0341.1}{10.1175/JCLI-D-19-0341.1})

\bibitem{risken}
Risken H. 1989 {\em {The Fokker-Planck Equation}}.
Springer second edition.

\bibitem{Bolley2012}
Bolley F, Gentil I, Guillin A. 2012  Convergence to equilibrium in Wasserstein
  distance for Fokker–Planck equations. {\em Journal of Functional Analysis}
  \textbf{263}, 2430--2457.
(\href{http://dx.doi.org/https://doi.org/10.1016/j.jfa.2012.07.007}{https://doi.org/10.1016/j.jfa.2012.07.007})

\bibitem{kloeden_1992}
Kloeden PE, Platen E. 1992 {\em Numerical solution of stochastic differential
  equations / Peter E. Kloeden, Eckhard Platen}.
Springer-Verlag Berlin ; New York.

\bibitem{gritsun2017}
Gritsun A, Lucarini V. 2017  Fluctuations, response, and resonances in a simple
  atmospheric model. {\em Physica D: Nonlinear Phenomena} \textbf{349}, 62--76.
(\href{http://dx.doi.org/10.1016/j.physd.2017.02.015}{10.1016/j.physd.2017.02.015})

\bibitem{Orcioni2014}
Orcioni S. 2014  Improving the approximation ability of {V}olterra series
  identified with a cross-correlation method. {\em Nonlinear Dynamics}
  \textbf{78}, 2861--2869.
(\href{http://dx.doi.org/10.1007/s11071-014-1631-7}{10.1007/s11071-014-1631-7})

\bibitem{Wray1994}
Wray J, Green GGR. 1994  Calculation of the {V}olterra kernels of non-linear
  dynamic systems using an artificial neural network. {\em Biological
  Cybernetics} \textbf{71}, 187--195.
(\href{http://dx.doi.org/10.1007/BF00202758}{10.1007/BF00202758})

\bibitem{Lucarini2020PRSA}
Lucarini V, Pavliotis GA, Zagli N. 2020  Response theory and phase transitions
  for the thermodynamic limit of interacting identical systems. {\em
  Proceedings of the Royal Society A: Mathematical, Physical and Engineering
  Sciences} \textbf{476}, 20200688.
(\href{http://dx.doi.org/10.1098/rspa.2020.0688}{10.1098/rspa.2020.0688})

\bibitem{Wang2013}
Wang Q. 2013  Forward and adjoint sensitivity computation of chaotic dynamical
  systems. {\em Journal of Computational Physics} \textbf{235}, 1 -- 13.
(\href{http://dx.doi.org/http://dx.doi.org/10.1016/j.jcp.2012.09.007}{http://dx.doi.org/10.1016/j.jcp.2012.09.007})

\bibitem{Chandramoorthy2020}
{Chandramoorthy} N, {Wang} Q. 2020  {A computable realization of Ruelle's
  formula for linear response of statistics in chaotic systems}. {\em arXiv
  e-prints} p. arXiv:2002.04117.

\bibitem{Ni2023}
Ni A. 2023  Fast Adjoint Algorithm for Linear Responses of Hyperbolic Chaos.
  {\em SIAM Journal on Applied Dynamical Systems} \textbf{22}, 2792--2824.
(\href{http://dx.doi.org/10.1137/22M1522383}{10.1137/22M1522383})

\bibitem{Ni2024}
Ni A, Tong Y. 2024  Equivariant Divergence Formula for Hyperbolic Chaotic
  Flows. {\em Journal of Statistical Physics} \textbf{191}.
(\href{http://dx.doi.org/10.1007/s10955-024-03329-1}{10.1007/s10955-024-03329-1})

\bibitem{Giorgini2024}
Giorgini LT, Deck K, Bischoff T, Souza A. 2024  Response Theory via Generative
  Score Modeling. {\em Phys. Rev. Lett.} \textbf{133}, 267302.
(\href{http://dx.doi.org/10.1103/PhysRevLett.133.267302}{10.1103/PhysRevLett.133.267302})

\bibitem{Ragone2016}
Ragone F, Lucarini V, Lunkeit F. 2016  A new framework for climate sensitivity
  and prediction: a modelling perspective. {\em Climate Dynamics} \textbf{46},
  1459--1471.
(\href{http://dx.doi.org/10.1007/s00382-015-2657-3}{10.1007/s00382-015-2657-3})

\bibitem{Lucarini2017}
Lucarini V, Ragone F, Lunkeit F. 2017  Predicting Climate Change Using Response
  Theory: Global Averages and Spatial Patterns. {\em Journal of Statistical
  Physics} \textbf{166}, 1036--1064.
(\href{http://dx.doi.org/10.1007/s10955-016-1506-z}{10.1007/s10955-016-1506-z})

\bibitem{Lembo2020}
Lembo V, Lucarini V, Ragone F. 2020  Beyond Forcing Scenarios: Predicting
  Climate Change through Response Operators in a Coupled General Circulation
  Model. {\em Scientific Reports} \textbf{10}, 8668.
(\href{http://dx.doi.org/10.1038/s41598-020-65297-2}{10.1038/s41598-020-65297-2})

\bibitem{MATLAB2024}
{The Mathworks, Inc.}. 2024 {\em {MATLAB version 24.1.0.2537033 (R2024a)}}.
Natick, Massachusetts.

\bibitem{JULIA}
Bezanson J, Edelman A, Karpinski S, Shah VB. 2017  Julia: A Fresh Approach to
  Numerical Computing. {\em SIAM Review} \textbf{59}, 65--98.
(\href{http://dx.doi.org/10.1137/141000671}{10.1137/141000671})

\bibitem{PYTHON}
{Python Software Foundation}. 2023 {\em {Python} 3.12.1 Documentation}.

\bibitem{OCTAVE}
Eaton JW, Bateman D, Hauberg S, Wehbring R. 2014 {\em {GNU Octave} version
  3.8.1 manual: a high-level interactive language for numerical computations}.
CreateSpace Independent Publishing Platform.

\bibitem{Tantet2018}
Tantet A, Lucarini V, Dijkstra HA. 2018a  {Resonances in a Chaotic Attractor
  Crisis of the Lorenz Flow}. {\em Journal of Statistical Physics}
  \textbf{170}, 584--616.
(\href{http://dx.doi.org/10.1007/s10955-017-1938-0}{10.1007/s10955-017-1938-0})

\bibitem{Tantet2018b}
Tantet A, Lucarini V, Lunkeit F, Dijkstra HA. 2018b  {Crisis of the chaotic
  attractor of a climate model: A transfer operator approach}. {\em
  Nonlinearity} \textbf{31}, 2221--2251.
(\href{http://dx.doi.org/10.1088/1361-6544/aaaf42}{10.1088/1361-6544/aaaf42})

\bibitem{Chek_al14_RP}
Chekroun MD, Neelin JD, Kondrashov D, McWilliams JC, Ghil M. 2014  {Rough
  parameter dependence in climate models: The role of Ruelle-Pollicott
  resonances}. {\em {Proc. Natl. Acad. Sci USA}} \textbf{111}, 1684--1690.
(\href{http://dx.doi.org/10.1073/pnas.1321816111}{10.1073/pnas.1321816111})

\bibitem{collet2013quasi}
Collet P, Mart{\'\i}nez S, San~Mart{\'\i}n J. 2013 {\em Quasi-Stationary
  Distributions: Markov Chains, Diffusions and Dynamical Systems}.
Probability and Its Applications. Springer.
(\href{http://dx.doi.org/10.1007/978-3-642-33131-2}{10.1007/978-3-642-33131-2})

\bibitem{Castro2024}
Castro MM, Lamb JS, Olicón-Méndez G, Rasmussen M. 2024  Existence and
  uniqueness of quasi-stationary and quasi-ergodic measures for absorbing
  Markov chains: A Banach lattice approach. {\em Stochastic Processes and their
  Applications} \textbf{173}, 104364.
(\href{http://dx.doi.org/https://doi.org/10.1016/j.spa.2024.104364}{https://doi.org/10.1016/j.spa.2024.104364})

\bibitem{Senne2012}
Senne M, Trendelkamp-Schroer B, Mey ASJS, Sch{\"u}tte C, No{\'e} F. 2012
  {EMMA}: A Software Package for Markov Model Building and Analysis. {\em
  Journal of Chemical Theory and Computation} \textbf{8}, 2223--2238.
(\href{http://dx.doi.org/10.1021/ct300274u}{10.1021/ct300274u})

\bibitem{Scherer2015}
Scherer MK, Trendelkamp-Schroer B, Paul F, P{\'e}rez-Hern{\'a}ndez G, Hoffmann
  M, Plattner N, Wehmeyer C, Prinz JH, No{\'e} F. 2015  {PyEMMA} 2: A Software
  Package for Estimation, Validation, and Analysis of Markov Models. {\em
  Journal of Chemical Theory and Computation} \textbf{11}, 5525--5542.
(\href{http://dx.doi.org/10.1021/acs.jctc.5b00743}{10.1021/acs.jctc.5b00743})

\bibitem{Harrigan2017}
Harrigan MP, Sultan MM, Hernández CX, Husic BE, Eastman P, Schwantes CR,
  Beauchamp KA, McGibbon RT, Pande VS. 2017  {MSMBuilder}: Statistical Models
  for Biomolecular Dynamics. {\em Biophysical Journal} \textbf{112}, 10--15.
(\href{http://dx.doi.org/10.1016/j.bpj.2016.10.042}{10.1016/j.bpj.2016.10.042})

\bibitem{Springeretal2024}
Springer S, Laio A, Galfi VM, Lucarini V. 2024  Unsupervised detection of
  large-scale weather patterns in the northern hemisphere via Markov State
  Modelling: from blockings to teleconnections. {\em npj Climate and
  Atmospheric Science} \textbf{7}, 105.
(\href{http://dx.doi.org/10.1038/s41612-024-00659-5}{10.1038/s41612-024-00659-5})

\bibitem{Suarez2021}
Suárez E, Wiewiora RP, Wehmeyer C, Noé F, Chodera JD, Zuckerman DM. 2021
  What Markov State Models Can and Cannot Do: Correlation versus Path-Based
  Observables in Protein-Folding Models. {\em Journal of Chemical Theory and
  Computation} \textbf{17}, 3119--3133.
(\href{http://dx.doi.org/10.1021/acs.jctc.0c01154}{10.1021/acs.jctc.0c01154})

\bibitem{Souza2024}
Souza AN, Silvestri S. 2024  A Modified Bisecting K-Means for Approximating
  Transfer Operators: Application to the Lorenz Equations. .

\bibitem{ColbrookTownsend2024}
Colbrook MJ, Townsend A. 2024  Rigorous data-driven computation of spectral
  properties of Koopman operators for dynamical systems. {\em Communications on
  Pure and Applied Mathematics} \textbf{77}, 221--283.
(\href{http://dx.doi.org/https://doi.org/10.1002/cpa.22125}{https://doi.org/10.1002/cpa.22125})

\bibitem{Wu2017}
Wu SJ, Chu MT. 2017  Markov chains with memory, tensor formulation, and the
  dynamics of power iteration. {\em Applied Mathematics and Computation}
  \textbf{303}, 226--239.
(\href{http://dx.doi.org/https://doi.org/10.1016/j.amc.2017.01.030}{https://doi.org/10.1016/j.amc.2017.01.030})

\bibitem{Mor2021}
Mor B, Garhwal S, Kumar A. 2021  A Systematic Review of Hidden Markov Models
  and Their Applications. {\em Archives of Computational Methods in
  Engineering} \textbf{28}, 1429--1448.
(\href{http://dx.doi.org/10.1007/s11831-020-09422-4}{10.1007/s11831-020-09422-4})

\bibitem{LucariniChekroun2024}
Lucarini V, Chekroun MD. 2024  Detecting and Attributing Change in Climate and
  Complex Systems: Foundations, Green's Functions, and Nonlinear Fingerprints.
  {\em Phys. Rev. Lett.} \textbf{133}, 244201.
(\href{http://dx.doi.org/10.1103/PhysRevLett.133.244201}{10.1103/PhysRevLett.133.244201})

\bibitem{Hasselmann1997}
Hasselmann K. 1997  Multi-pattern fingerprint method for detection and
  attribution of climate change. {\em Climate Dynamics} \textbf{13}, 601--611.
(\href{http://dx.doi.org/10.1007/s003820050185}{10.1007/s003820050185})

\bibitem{Allen1999}
Allen M, Tett S. 1999  Checking for model consistency in optimal
  fingerprinting. {\em Climate Dynamics} \textbf{15}, 419--434.
(\href{http://dx.doi.org/10.1007/s003820050291}{10.1007/s003820050291})

\bibitem{Hegerl2011}
Hegerl G, Zwiers F. 2011  Use of models in detection and attribution of climate
  change. {\em WIREs Climate Change} \textbf{2}, 570--591.
(\href{http://dx.doi.org/https://doi.org/10.1002/wcc.121}{https://doi.org/10.1002/wcc.121})

\bibitem{Hannart2014}
Hannart A, Ribes A, Naveau P. 2014  Optimal fingerprinting under multiple
  sources of uncertainty. {\em Geophysical Research Letters} \textbf{41},
  1261--1268.
(\href{http://dx.doi.org/https://doi.org/10.1002/2013GL058653}{https://doi.org/10.1002/2013GL058653})

\bibitem{Susuki2015}
Susuki Y, Mezić I. 2015  A prony approximation of Koopman Mode Decomposition.
  In {\em 2015 54th IEEE Conference on Decision and Control (CDC)} pp.
  7022--7027.
(\href{http://dx.doi.org/10.1109/CDC.2015.7403326}{10.1109/CDC.2015.7403326})

\end{thebibliography}
\end{document}